\documentclass[twocolumn,10pt]{article}
\usepackage{amsmath}
\usepackage{amssymb}
\usepackage{graphicx}
\usepackage{cite}
\usepackage{color} 
\usepackage{authblk}
\usepackage{subfig}
\newcommand{\gene}[1]{\textit{#1}}

\begin{document}

\title{Genome-wide analysis points to roles for extracellular matrix remodeling, 
the visual cycle, and neuronal development in myopia}

\author[1]{Amy K. Kiefer}
\author[1]{Joyce Y. Tung}
\author[1]{Chuong B. Do}
\author[1]{David A. Hinds}
\author[1]{Joanna L. Mountain}
\author[1]{Uta Francke}
\author[1,*]{Nicholas Eriksson}
\affil[1]{23andMe, Inc., Mountain View, CA, USA}
\affil[*]{\texttt{nick@23andme.com}}
\maketitle

\begin{abstract}

Myopia, or nearsightedness, is the most common eye disorder, resulting
primarily from excess elongation of the eye.
The etiology of myopia, although known to be complex, is
poorly understood. Here we report the largest ever genome-wide association
study (43,360 participants) on myopia in Europeans. We performed a survival
analysis on age of myopia onset and identified 19 significant associations ($p < 5\cdot
10^{-8}$), two of which are replications of earlier associations with
refractive error. These 19 associations in total explain 2.7\% of the variance 
in myopia age of onset, and point towards a number of different mechanisms behind 
the development of myopia. One association is in the gene 
\gene{PRSS56}, which has previously been linked to
abnormally small eyes; one is in a gene that forms part of the
extracellular matrix (\gene{LAMA2}); two are in or near genes involved in
the regeneration of 11-cis-retinal (\gene{RGR} and
\gene{RDH5}); two are near genes known to be involved in the growth and
guidance of retinal ganglion cells (\gene{ZIC2}, \gene{SFRP1}); and five are in or near
genes involved in neuronal signaling or development. These novel
findings point towards multiple genetic factors involved in the development of
myopia and suggest that complex interactions between extracellular matrix
remodeling, neuronal development, and visual signals from the retina may underlie
the development of myopia in humans. 
\end{abstract}

\section*{Author Summary}
The genetic basis of myopia, or nearsightedness, is believed to be complex and
affected by multiple genes. Two genetic association studies have each
identified a single genetic region associated with myopia in European populations. Here we report the
results of the largest ever genetic association study on myopia in over 40,000
people of European ancestry. We identified 19 genetic regions significantly associated with myopia
age of onset.  Two are replications of the previously identified associations,
and 17 are novel. Thirteen of the novel associations are in or near genes
implicated in eye development, neuronal development and signaling, the visual 
cycle of the retina, and general morphology: \gene{DLG2}, \gene{KCNMA1}, \gene{KCNQ5},
\gene{LAMA2}, \gene{LRRC4C}, \gene{PRSS56}, \gene{RBFOX1}, \gene{RDH5},
\gene{RGR}, \gene{SFRP1}, \gene{TJP2}, \gene{ZBTB38}, and \gene{ZIC2}. These
findings point to numerous biological pathways involved in the development of
myopia, and in particular, suggest that early eye and neuronal development may
lead to the eventual development of myopia in humans.

\section*{Introduction}
Myopia, or nearsightedness, is the most common eye disorder worldwide. In the
United States, an estimated 30-40\% of the adult population has clinically
relevant myopia (more severe than -1 diopter), and the prevalence has increased
markedly in the last 30 years \cite{pmid20008719, pmid15078666}. Myopia is a
refractive error that results primarily from increased axial length of the eye
\cite{pmid21155761}. The increased physical length of the eye relative to
optical length causes images to be focused in front of the retina, resulting in
blurred vision. 

The etiology of myopia is multifactorial \cite{pmid21155761}. Briefly, postnatal eye
growth is directed by visual stimuli that evoke a signaling cascade within the
eye.  This cascade is initiated in the retina and passes through the choroid to
 guide remodeling of the sclera (the white part of the eye) (cf.\
\cite{pmid16079001, pmid16202407}). Animal models implicate these
visually-guided alterations of the scleral extracellular matrix in the eventual
development of myopia. \cite{pmid7610579, pmid16079001}.

In humans the eye typically grows about 5 mm from birth to age six, during
which time vision typically improves \cite{pmid8631638}. At age six only about
2\% of children are myopic \cite{pmid8631638}. Although the eye grows only 0.5
mm through puberty \cite{pmid7136552}, the incidence of myopia increases
sevenfold \cite{pmid8631638}, peaking between the ages 9--14 \cite{pmid15037570}.
Myopia developed during childhood or early adolescence generally worsens
throughout adolescence and then stabilizes by age 20. Compared to myopia that
develops in childhood or adolescence, adult onset myopia tends to be less
severe \cite{pmid8751115, pmid19353396, pmid14977519}.

Although epidemiological studies have implicated numerous environmental factors
in the development of myopia, most notably education, outdoor exposure,
reading, and near work \cite{pmid21155761}, it is well established that
genetics plays a substantial role. Twin and sibling studies have provided heritability
estimates that range from 50\% to over 90\% \cite{pmid11328732, pmid11734523,
pmid1937488, pmid15851555, pmid17724179}. Children of myopic parents tend to
have longer eyes and are at increased risk of developing myopia in childhood
\cite{pmid8728496}. Segregation analyses suggest that multiple genes are
involved in the development of myopia \cite{pmid15671267, pmid4029963}. To
date, there have been seven genome-wide association studies (GWAS) on myopia or
related phenotypes (pathological myopia, refractive index, and ocular axial
length): two in Europeans \cite{pmid20835239, pmid20835236} and five
in Asian populations \cite{pmid21095009, pmid19779542, pmid21505071,
pmid21640322, pmid22685421}. Each of these publications has
identified a different single association with myopia. In addition there have
been several linkage studies (see \cite{pmid17210850,pmid21155761} for reviews) and an
exome sequencing study of severe myopia \cite{pmid21695231}.  

In contrast to the previous relatively small (up to approximately 5,000 initial
cases) GWAS that used degree of refractive error as a quantitative dependent
measure, we analyzed data for 43,360 individuals from the 23andMe database who
reported whether they had been diagnosed with nearsightedness, and if so, at
what age. We performed a genome-wide survival analysis on age of onset of
myopia, discovering 19 genome-wide significant associations with myopia age of
onset, 17 of which are novel.

\section*{Results and Discussion}

Participants reported via a web-based questionnaire whether they had been
diagnosed with nearsightedness, and if so, at what age. All participants were
customers of 23andMe and of primarily European ancestry; no pair was more
closely related than at the level of first cousins. We performed a genome-wide
survival analysis using a Cox proportional hazards model on 43,360 individuals
(``discovery set''). This model assumes that there is an (unknown) baseline
probability of developing myopia at every year of age. The model then tests whether each
single nucleotide polymorphism (SNP) is associated with a significantly higher or lower 
probability of developing myopia compared to baseline. The Cox model can be thought of as a
generalization of an analysis of myopia age of onset. In contrast to an analysis of age of 
onset, the Cox model allows for the inclusion of
non-myopic controls, resulting in considerably increased statistical power. Analyses controlled for sex and five principal
components of genetic ancestry. An additional, non-overlapping set of 4,277
participants who answered a separate question about their use of corrective eyewear for
nearsightedness before the age of ten were used as a
replication set. See Table~\ref{tab:tab1} for characteristics of the two
cohorts.

\begin{table*}
\centering
\begin{tabular}{ccccc}
\hline\hline
                              & Number &\% female   & Age (SE)      & Age of onset (SE) \\ \hline
Discovery, myopic             & 26038  &     46.1   &   48.6 (15.7) &     16.4 (11.0)   \\
Discovery, not myopic         & 17322  &     39.6   &   49.1 (17.1) &      ---          \\ \hline
Replication, myopic at 10     &   800  &     45.1   &   47.7 (14.9) &     $\leq10$      \\
Replication, not myopic at 10 &  3477  &     45.2   &   50.0 (16.6) &      ---          \\
\hline\hline
\end{tabular}
\caption{
\textbf{Cohort statistics.}
Sex, current age, and age of onset for discovery and replication cohorts. 
}
\label{tab:tab1}
\end{table*}

\begin{figure*}
\begin{center}
\includegraphics[width=.99\textwidth]{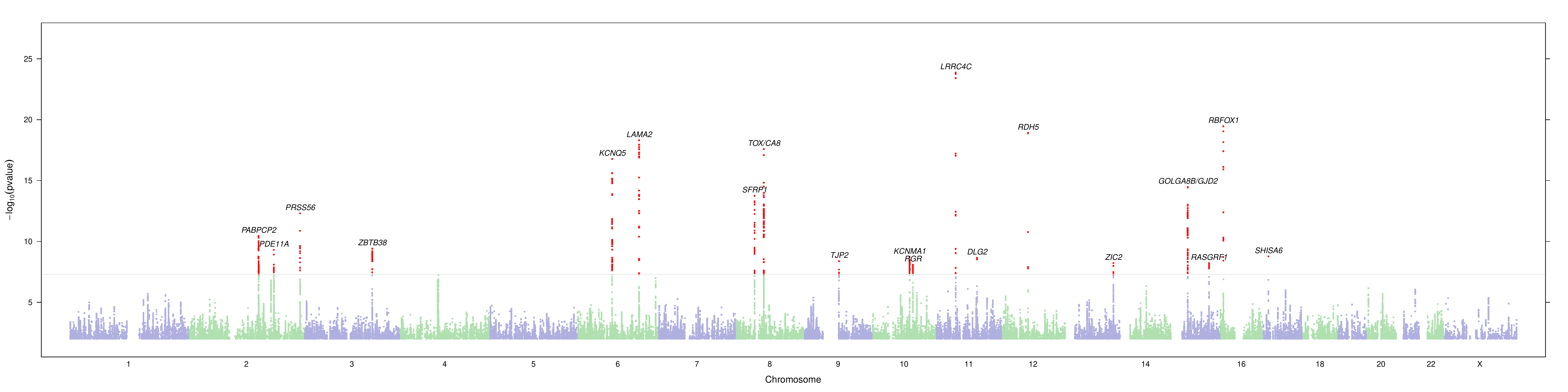}
\end{center}
\caption{
{\bf Negative $\log_{10} p$-values genome wide for myopia.}  
Regions are named with their postulated candidate gene or genes.
$p$-values under $10^{-25}$ have been cutoff (only the \gene{LAMA2} region is affected).
See Figure~\ref{fig:regions} for plots in each region with a significant association.  }
\label{fig:manhattan}
\end{figure*}

Table~\ref{tab:tab2} shows the top SNPs for
all 27 genetic regions associated with myopia with a $p$-value smaller than
$10^{-6}$. All $p$-values from the GWAS have been corrected for the inflation
factor of 1.14. A total of 19 of the SNPs cross our threshold for genome-wide
significance ($5\cdot10^{-8}$, see Figure~\ref{fig:regions}). These 19 include
 two SNPs previously associated with refractive error in GWAS of European populations: rs524952
near \gene{GJD2} and \gene{ACTC1} and rs4778882  near
\gene{RASGRF1} \cite{pmid20835236,
pmid22665138, pmid20835239}. $p$-values genome-wide are shown in
Figure~\ref{fig:manhattan}; Figure~\ref{fig:qqplot} shows the quantile-quantile
plot for the analysis.

\begin{table*}
\centering
{\scriptsize
\begin{tabular}{cccccccccc}
\hline\hline
         rsid  & chr &   Position  &       Genes  &   MAF  &   $r^2$&allele &  HR    (CI)            &   $p$-value           & $p_{\rm{repl}}$ \\\hline
    rs12193446 &   6 &  129820038  &        \gene{LAMA2} & 0.094  & 0.991  &   A/G & 0.798 (0.773 -- 0.823) &   $  1\cdot 10^{-42}$ &   $4.9\cdot 10^{-4}$    \\
    rs11602008 &  11 &   40149305  &       \gene{LRRC4C} & 0.160  & 0.887  &   A/T & 1.149 (1.121 -- 1.177) &   $1.3\cdot 10^{-24}$ &   0.012  \\
    rs17648524 &  16 &    7459683  &       \gene{RBFOX1} & 0.364  & 0.974  &   G/C & 1.095 (1.075 -- 1.114) &   $3.5\cdot 10^{-20}$ &   0.27   \\
     rs3138142 &  12 &   56115585  &         \gene{RDH5} & 0.218  & 0.817  &   C/T & 0.892 (0.872 -- 0.913) &   $1.2\cdot 10^{-19}$ &   0.0074 \\
 chr8:60178580 &   8 &   60178580  &      \gene{TOX/CA8} & 0.358  & 0.970  &   C/G & 0.917 (0.900 -- 0.934) &   $2.6\cdot 10^{-18}$ &   0.26   \\
     rs7744813 &   6 &   73643289  &        \gene{KCNQ5} & 0.405  & 0.955  &   A/C & 0.920 (0.904 -- 0.937) &   $1.7\cdot 10^{-17}$ &   0.0016 \\
      rs524952 &  15 &   35005886  & \gene{GOLGA8B/GJD2} & 0.468  & 0.980  &   T/A & 1.078 (1.059 -- 1.097) &   $3.3\cdot 10^{-15}$ &   0.0019 \\
     rs2137277 &   8 &   40734662  &        \gene{SFRP1} & 0.189  & 0.922  &   A/G & 0.908 (0.887 -- 0.929) &   $1.8\cdot 10^{-14}$ &   0.52  \\
     rs1550094 &   2 &  233385396  &       \gene{PRSS56} & 0.306  & 0.963  &   A/G & 1.077 (1.057 -- 1.098) &   $4.9\cdot 10^{-13}$ &   0.019 \\
    rs11681122 &   2 &  146786063  &      \gene{PABPCP2} & 0.425  & 0.940  &   A/G & 0.937 (0.920 -- 0.954) &   $3.6\cdot 10^{-11}$ &   0.085 \\
     rs7624084 &   3 &  141093285  &       \gene{ZBTB38} & 0.435  & 0.961  &   T/C & 0.942 (0.925 -- 0.959) &   $3.8\cdot 10^{-10}$ &   0.19  \\
     rs1898585 &   2 &  178660450  &       \gene{PDE11A} & 0.163  & 0.942  &   C/T & 1.085 (1.059 -- 1.111) &   $4.9\cdot 10^{-10}$ &   0.011 \\
     rs2908972 &  17 &   11407259  &       \gene{SHISA6} & 0.397  & 0.970  &   T/A & 1.060 (1.042 -- 1.079) &   $1.7\cdot 10^{-9}$  &   0.053 \\
     rs6480859 &  10 &   79081948  &       \gene{KCNMA1} & 0.363  & 0.987  &   C/T & 1.060 (1.042 -- 1.079) &   $2.0\cdot 10^{-9}$  &   0.82  \\
    rs10736767 &  11 &   84637065  &         \gene{DLG2} & 0.451  & 0.996  &   A/C & 1.058 (1.040 -- 1.077) &   $2.2\cdot 10^{-9}$  &   0.53  \\
    rs11145746 &   9 &   71834380  &         \gene{TJP2} & 0.198  & 0.886  &   G/A & 1.076 (1.052 -- 1.100) &   $4.2\cdot 10^{-9}$  &   0.77  \\
     rs4291789 &  13 &  100672921  &         \gene{ZIC2} & 0.326  & 0.724  &   C/G & 1.070 (1.048 -- 1.093) &   $  6\cdot 10^{-9}$  &   $2.2\cdot 10^{-4}$ \\
     rs4778882 &  15 &   79382019  &      \gene{RASGRF1} & 0.399  & 0.951  &   A/G & 1.059 (1.040 -- 1.078) &   $6.1\cdot 10^{-9}$  &   0.017 \\
      rs745480 &  10 &   85986554  &          \gene{RGR} & 0.474  & 0.975  &   C/G & 1.056 (1.038 -- 1.075) &   $  8\cdot 10^{-9}$  &   0.32  \\ \hline
     rs5022942 &   4 &   81959966  &         \gene{BMP3} & 0.229  & 0.991  &   G/A & 1.063 (1.041 -- 1.085) &   $5.9\cdot 10^{-8}$  &   0.21  \\
     rs9365619 &   6 &  164251746  &          \gene{QKI} & 0.457  & 0.999  &   C/A & 1.051 (1.033 -- 1.069) &   $  1\cdot 10^{-7}$  &   0.097 \\
     rs1031004 &   4 &   80516849  &       \gene{ANTXR2} & 0.261  & 0.983  &   T/A & 1.058 (1.037 -- 1.079) &   $1.5\cdot 10^{-7}$  &   0.62  \\
    rs17428076 &   2 &  172851936  &   \gene{HAT1/MAP1D} & 0.236  & 0.985  &   C/G & 0.943 (0.924 -- 0.963) &   $1.6\cdot 10^{-7}$  &   0.18  \\
chr14:54413001 &  14 &   54413001  &         \gene{BMP4} & 0.489  & 0.930  &   G/C & 0.952 (0.935 -- 0.969) &   $4.6\cdot 10^{-7}$  &   0.38  \\
     rs7272323 &  20 &    4756691  &  \gene{PRND/RASSF2} & 0.409  & 0.955  &   C/G & 1.050 (1.031 -- 1.068) &   $  7\cdot 10^{-7}$  &   0.015 \\
chr11:65348347 &  11 &   65348347  &      \gene{EHBP1L1} & 0.017  & 0.558  &   G/A & 0.781 (0.711 -- 0.858) &   $7.9\cdot 10^{-7}$  &   0.97  \\
    rs55819392 &  21 &   40038207  &     \gene{ERG/ETS2} & 0.259  & 0.987  &   G/A & 0.949 (0.930 -- 0.968) &   $9.2\cdot 10^{-7}$  &   0.014 \\
\hline\hline
\end{tabular}
}
\caption{
\textbf{Index SNPs for regions with $p<10^{-6}$.}
Index SNPs for regions with ($\lambda$-corrected) $p$-values under $10^{-6}$.
Positions and alleles are given relative to the positive strand of build 37 of
the human genome; alleles are listed as major/minor.  The listed genes are the
postulated candidate gene in each region.  $r^2$ is the estimated imputation
accuracy; HR is the hazard ratio per copy of the minor allele; $p$-value is the $p$-value
in the discovery cohort; $p_{\rm{repl}}$ is the $p$-value in the replication cohort.
}
\label{tab:tab2}
\end{table*}

\begin{figure}
\begin{center}
\includegraphics[width=.5\textwidth]{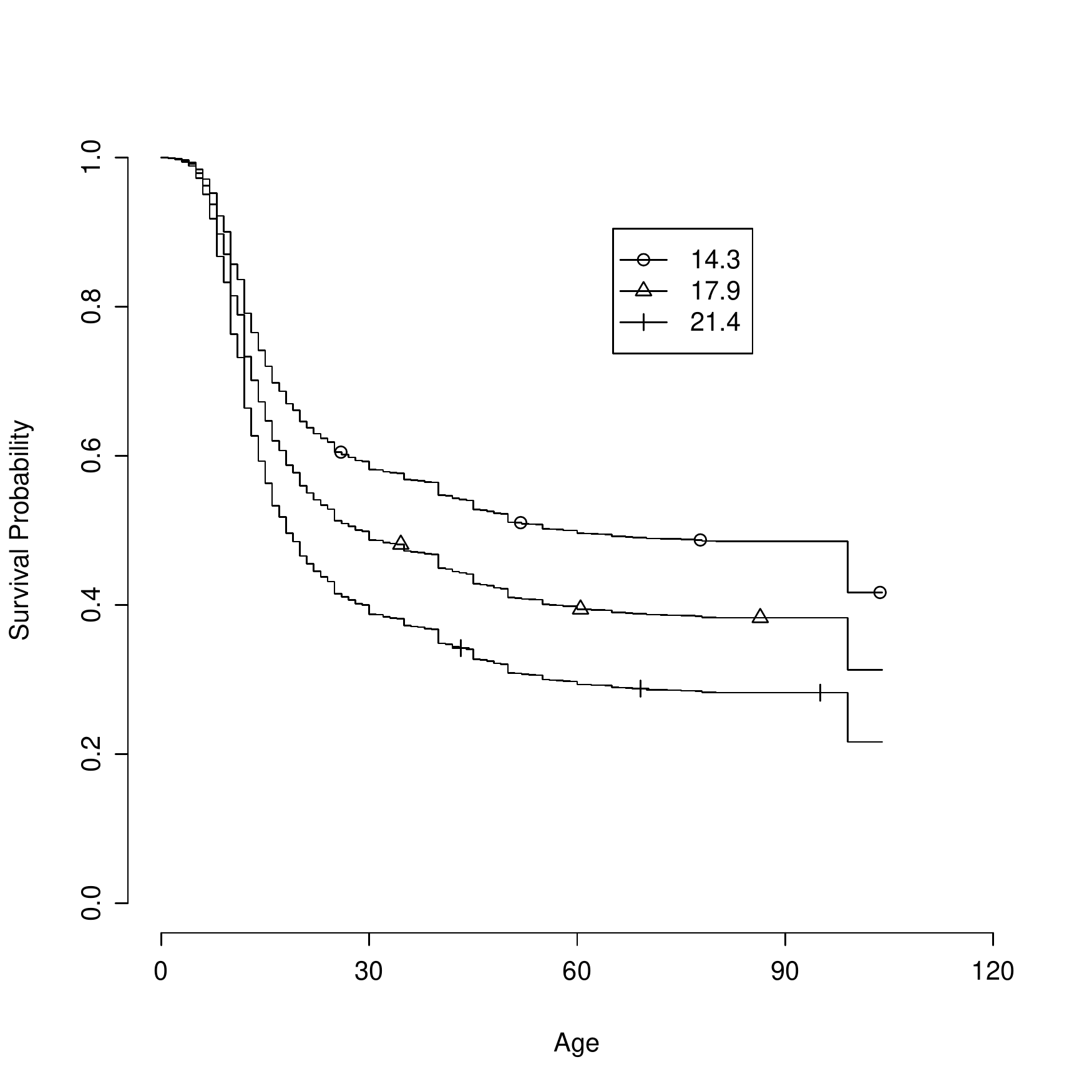}
\end{center}
\caption{
\textbf{Estimated survival curves by genetic propensity score.} 
The genetic propensity score is computed as the number of risk alleles across the 19
genome-wide significant SNPs.  Curves show estimated survival probability (i.e., the probability of not having developed myopia) by age under the fitted Cox model for the 10th, 50th, and 90th percentiles
of scores (14.3, 17.9, and 21.4, respectively).  }
\label{fig:survival}
\end{figure}

Of the 19 SNPs significant in the discovery set, nine were also significant in
the replication set (Table~\ref{tab:tab2}). As the replication set was small
(barely a tenth the size of the discovery set) and measured age of onset less
exactly, it is not surprising that not all SNPs replicated. We defined a
genetic myopia propensity score as the number of copies of the risk alleles
across all 19 SNPs identified via the discovery set. The propensity score
showed a strong association with early onset myopia (less than 10 years old) in
our replication cohort ($p=6.5\cdot10^{-12}$, odds ratio 1.08 per risk allele).
The top decile of genetic propensity had 2.54 greater odds of developing myopia
before the age of 10 than the bottom decile. In a Cox model fit
to the discovery set, the propensity score explains 2.7\% of the total variance.
Note that this estimate may be inflated, as it
is calculated on the discovery population.  
In this model, someone in the 90th percentile of risk (a score of 21.4) is
nearly twice as likely to develop myopia by the age of 60 as someone in the
10th percentile of risk (score of 14.3), Figure~\ref{fig:survival}.

Of the 17 novel associations, many lie in or near genes with direct links to
processes related to myopia development. Below, we briefly sketch out possible
connections between these associations and extracellular matrix (ECM)
remodeling, the visual cycle, eye and body growth, retinal neuron
development, and general neuronal development or signaling.

\subsection*{Extracellular Matrix Remodeling}

The strongest association is a SNP in an intron of \gene{LAMA2} (laminin, alpha
2 subunit, rs12193446, $p=1.0\cdot 10^{-42}$, hazard ratio (HR) 0.80).
Laminins are extracellular structural proteins that are integral parts of the
ECM. Changes in the composition of the ECM of the sclera have been shown to
alter the axial length of the eye \cite{pmid16202407}. Laminins play a role in the
development and maintenance of different eye structures \cite{pmid16505007,
pmid7649817}. The laminin alpha 2 chain is found in the extraocular muscles
during development \cite{pmid16505007}, and may act as an adhesive
substrate and possibly a guidance cue for retinal ganglion cell growth cones
 \cite{pmid8613743}. We also found a suggestive
association related to laminin (rs1031004, $p=1.5\cdot 10^{-7}$, HR=1.06) 312
kb upstream of \gene{ANTXR2} (anthrax toxin receptor 2). ANTXR2 binds laminin
and possibly collagen type IV \cite{pmid11683410} and thus may also be involved
in extracellular matrix remodeling.

\subsection*{The Visual Cycle}

Two of the novel associations are in or near genes involved in the regeneration of 
11-cis-retinal, the light sensitive component of photoreceptors; this regeneration is
commonly referred to as the visual cycle of the retina. 
These associations are with rs3138142, $p=1.2\cdot10^{-19}$, HR=0.89 in
\gene{RDH5} (retinol dehydrogenase 5 (11-cis/9-cis)) and rs745480
($p=8.0\cdot10^{-9}$, HR=1.06), a SNP 18 kb upstream of \gene{RGR}, which encodes the
retinal G protein-coupled receptor. 
The SNP rs3138142 is a synonymous change in \gene{RDH5}.  It  has been linked
to \gene{RDH5} expression \cite{pmid18846210,pmid17873874}, and
it is part of an Nr2f2 (nuclear receptor subfamily 2, group F, member 2)
transcription factor binding motif in mouse \cite{pmid19443739, regulomedb}.
Both \gene{RDH5} and \gene{RGR} play crucial roles in the
regeneration of 11-cis retinal in the retinal pigment epithelium (RPE) \cite{pmid15987797}.
Mutations in \gene{RDH5} have been linked with fundus albipunctatus, a rare form of
congenital stationary night blindness (for a recent review, see
\cite{pmid22669287}) and progressive cone dystrophy \cite{pmid11053295}, and
mutations in \gene{RGR} have been linked with autosomal recessive and autosomal
dominant retinitis pigmentosa \cite{pmid11559856, pmid12843338}. 

We also identified an association within another gene that functions in the RPE:
rs7744813 ($p=1.7\cdot 10^{-17}$, HR=0.92), a SNP in \gene{KCNQ5} (potassium
voltage-gated channel, KQT-like subfamily, member 5). \gene{KCNQ5} encodes a potassium channel
 found in the RPE and neural retina. These channels are believed to contribute
to ion flow across the RPE \cite{pmid22135213, pmid21795522} and to affect the
function of cone and rod photoreceptors \cite{pmid21795522}.

\subsection*{Eye and Body Growth}
Three of our associations show possible links to eye or overall morphology.
The first is a missense mutation in \gene{PRSS56} (A224T, rs1550094,
$p=4.9\cdot 10^{-13}$, HR=1.08). Other mutations in \gene{PRSS56} have been
shown to cause strikingly small eyes with severe decreases in axial length
\cite{pmid21532570, pmid21850159, pmid21397065}. Two other associated SNPs are
in linkage disequilibrium with SNPs previously associated with height:
rs10113215 ($p=2.6\cdot 10^{-18}$, HR=0.92), near \gene{TOX} and \gene{CA8}
(thymus high mobility group box protein; carbonic anhydrase VIII), and
rs7624084 ($p=3.8\cdot 10^{-10}$, HR=0.94), near \gene{ZBTB38} (zinc finger and
BTB domain-containing protein 38). The SNPs rs10113215 and rs7624084 are in
linkage disequilibrium (LD) with rs6569648 and rs6763931, respectively ($r^2 >
0.6$ and $r^2 > 0.8$); both of which have been associated with height
\cite{pmid20881960,pmid18391951}.

\subsection*{Retinal Ganglion Cell Projections}
Two of the novel associations are near genes that affect the outgrowth of
retinal ganglion neurons during development. The first is rs4291789
($p=6.0\cdot10^{-9}$, HR=1.07), which lies 34 kb downstream of \gene{ZIC2} (Zic
family member 2). ZIC2 regulates two independent parts of ipsilateral retinal
ganglion cell development: axon repulsion at the optic chiasm midline
\cite{pmid18417618, pmid13678579}, and organization of the axonal projections 
at their final targets in the brain \cite{pmid20676059}. 

The second, rs2137277 ($p=1.8\cdot10^{-14}$, HR=0.91), is a variant in
\gene{ZMAT4} (zinc finger, matrin-type 4). \gene{ZMAT4} has no known link to
vision, but this variant also lies 385 kb downstream of \gene{SFRP1} (secreted
frizzled-related protein 1). \gene{SFRP1} is involved in the differentiation
of the optic cup from the neural retina \cite{pmid21896628}, retinal
neurogenesis \cite{pmid21478884}, the development and function of photoreceptor
cells \cite{pmid15235574, pmid12724355}, and the growth of retinal ganglion
cells \cite{pmid16172602}.

\subsection*{Neuronal Signaling and Development}

Finally, we found five associations with SNPs in genes involved in
neuronal development and signaling, but without a known role in vision
development or the vision cycle: in \gene{KCNMA1} (potassium large conductance
calcium-activated channel, subfamily M, alpha member 1; rs6480859,
$p=2.0\cdot10^{-9}$, HR=1.06); in \gene{RBFOX1} (RNA binding protein, fox-1
homolog;  rs17648524, $p=3.5\cdot10^{-20}$, HR=1.10); in \gene{LRRC4C},
leucine rich repeating region containing 4C, also known as \gene{NGL-1} (rs11602008,
$p=1.3\cdot10^{-24}$, HR=1.15); in \gene{DLG2} (discs, large homolog 2;
rs10736767, $p=2.2\cdot10^{-9}$, HR=1.06); and in \gene{TJP2} (tight junction
protein 2; rs11145746, $p=4.2\cdot10^{-9}$, HR=1.08). 

\gene{KCNMA1} encodes the pore-forming alpha subunit of a MaxiK channel, a family 
of large conductance, voltage and calcium-sensitive potassium channels involved 
in the control of smooth muscle and neuronal excitation. \gene{RBFOX1} belongs to a
family of RNA binding proteins that regulates the alternative splicing of
several neuronal transcripts implicated in neuronal development and maturation
\cite{pmid22730494}. \gene{LRRC4C} encodes a binding partner for netrin G1 and 
promotes the outgrowth of thalamocortical axons \cite{pmid14595443}. \gene{DLG2} 
plays a critical role in the formation and regulation of protein scaffolding 
at postsynaptic sites \cite{pmid21739617}. \gene{TJP2} has been linked with hearing loss: 
its duplication and subsequent overexpression are found in adult-onset progressive 
nonsyndromic hearing loss \cite{pmid20602916}. 

\subsection*{Conclusion}

This study represents the largest GWAS on myopia in Europeans to date.
This cohort of 43,360 individuals led to the discovery of 17 novel associations as
well as replication of the two previously reported associations in Europeans.
In contrast to the earlier studies that used refractive error as a quantitative 
outcome, we used a Cox proportional hazards model with age of onset of myopia as our major endpoint. 
This model yielded greater statistical power than 
a simple case-control study of myopia. Of the 19 significant SNPs found using this model,
all but
one had smaller $p$-values when a hazards model was employed, and 
only 13 would be genome-wide significant using a
case-control analysis on the same dataset (Table~\ref{tab:sup1}).

The proportional hazards model assumes that each SNP has a constant effect on
the hazard of developing myopia at any age. When we tested the validity of this
assumption for the 19 significant SNPs, five showed evidence of
different effects at different ages (Table~\ref{tab:sup_prop}). While this violation should not
lead to overly small $p$-values for these SNPs in the GWAS, it does make risk
prediction based on these results less straightforward.  
These age dependent hazards suggest that different biological processes may
affect the development of myopia at different ages. For example, rs12193446 in
\gene{LAMA2} shows a large effect on myopia hazard at an early age, peaking
around 11 years, and then a null or even negative effect on hazard after the
age of 30; other SNPs show different patterns of effect as a
function of age (Figure~\ref{fig:proportional}).

Our findings further suggest that there may be somewhat different genetic factors
underlying myopia age of onset and refractive error. Because adult onset myopia tends
to be less severe than myopia developed in childhood or adolescence \cite{pmid8751115, pmid19353396, pmid14977519}, age of onset is likely
correlated with refractive error, but it is not known how strongly.  Many of our
associations showed a stronger signal than the two associations near
\gene{GJD2} and near \gene{RASGRF1} previously linked with refractive error in
Europeans. Notably, the latter association, near \gene{RASGRF1}, also failed to
replicate in a recent meta-analysis  \cite{pmid22665138}.
The fact that many of our associations with strong effects on age of onset
have not shown up in previous refractive error GWAS implies that
some genetic factors may affect the age of
onset independent of eventual severity, and that the strength of different genetic
associations with myopia may depend on the specific phenotype under study. 

We also note that our phenotype was based on participants' reports rather than
clinical assessments. Although in theory errors in recall could have
affected our results, we expect that the vast majority of people are able to recall
when they first wore glasses with at most a few years of error. 

The five associations previously reported in pathological myopia or
refractive error GWAS in Asian populations \cite{pmid21095009, pmid19779542, pmid21505071,
pmid21640322, pmid22685421}  show no overlap with the significant
or suggestive regions found here. Nor did we find an association with the
\gene{ZNF644} locus that was identified as the site of high-penetrance,
autosomal dominant mutations in Han Chinese families with apparent monogenic
inheritance of high-grade myopia \cite{pmid21695231}. This lack of overlap is
further evidence that the genetic factors involved in myopia differ across populations.

Our identification of 17 novel genetic associations suggests several novel genetic pathways 
in the development of human myopia. These findings augment existing research 
on the development of myopia, which to date has been studied primarily in animal models 
of artificially induced myopia. Some of the associations are consistent with 
the current view, based largely on animal models, that a visually-triggered signaling cascade from the retina 
ultimately guides the scleral remodeling that leads to eye growth, and 
that the RPE plays a key role in
this process \cite{pmid16079001}. A number of the novel associations 
point to the potential importance of early neuronal development in the eventual development of 
myopia, particularly the growth and topographical organization of retinal ganglion cells.
These associations suggest that early neuronal development may also contribute to future refractive
errors. We expect that these findings will drive new research into the complex
etiology of myopia. 

\section*{Methods}

\subsection*{Human Subjects}
All participants were drawn from the customer base of 23andMe, Inc., a consumer
genetics company. This cohort has been described in detail previously
\cite{pmid20585627, pmid21858135}. Participants provided informed consent and
participated in the research online, under a protocol approved by the external
AAHRPP-accredited IRB, Ethical \& Independent Review Services (E\&I Review).

\subsection*{Phenotype data}
Participants in the discovery cohort were asked online as part of a medical
history questionnaire: ``Have you ever been diagnosed by a doctor with any of the
following vision conditions?: Nearsightedness (near objects are clear, far
objects are blurry) (Yes/No/I don't know)''. If they answered ``yes'', they were
asked, ``At what age were you first diagnosed with nearsightedness (near objects
are clear, far objects are blurry)? Your best guess is fine.'' Those reporting
an age of onset outside of the range 0--100 were removed from analysis. All
participants also reported current age.

The replication cohort consisted of 23andMe customers who were not part of the
discovery cohort (i.e., they didn't provide a nearsightedness diagnosis). They
answered a single question ``Did you wear glasses or other corrective eyewear
for nearsightedness before the age of 10? (Yes/No/I'm not sure)''.

\subsection*{Genotyping and imputation}
Participants were genotyped and additional SNP genotypes were imputed against
the August 2010 release of the 1000 genomes data as described
previously \cite{pmid22747683}. Briefly, they were genotyped on at least one of
three genotyping platforms, two based on the Illumina HumanHap550+ BeadChip,
the third based on the Illumina Human OmniExpress+ BeadChip. The platforms included
assays for 586,916, 584,942, and 1,008,948 SNPs respectively. Genotypes for a
total of 11,914,767 SNPs were imputed in batches of roughly 10,000 individuals,
grouped by genotyping platform. Of these, 7,087,609 met our criteria of 0.005 minor
allele frequency, average $r^2$ across batches of at least 0.5, and minimum
$r^2$ across batches of at least 0.3. (The minimum $r^2$ requirement was added
to filter out SNPs that imputed poorly in the batches consisting of the less
dense platform.)

\subsection*{Statistical analysis}
In order to minimize population substructure while maximizing statistical
power, the study was limited to individuals with European ancestry. Ancestry
was inferred from the genome-wide genotype data, and principal component
analysis was performed as in \cite{pmid22493691,pmid20585627}. No two
participants shared more than 700 cM of DNA identical by descent (IBD,
approximately the lower end of sharing between a pair of first cousins). IBD
was calculated using the methods described in \cite{pmid22509285}.

For the survival analysis, let the hazard function $h(t)$ be the rate of
developing myopia at time $t$. Then the Cox proportional hazards model is 
\[ \log h(t) = \alpha(t) + \beta_g G + \beta_s S + \sum_{i=1}^{5} \beta_{pc_i} PC_i \]
for an arbitrary baseline hazard function $\alpha(t)$ and covariates $G$
(genotype), $S$ (sex) and $PC_1, \dots, PC_5$ (projections onto principal
components). $G$ was coded as a dosage from 0--2 as the estimated number of
minor alleles present.

For each SNP, we fit a Cox proportional hazards model using R \cite{Rsurv} and computed a
p-value using a likelihood ratio test for the genotype term. All SNPs with
$p$-values under $5\cdot 10^{-8}$ after genomic control correction were
considered genome-wide significant. The hazard ratio (HR) reported throughout
can be interpreted as the multiplicative change in the rate of onset of myopia
per copy of the minor allele (e.g., $e^{\beta_g}$). 
To test the proportional hazards assumption, we tested for independence of the
scaled Schoenfeld residuals for each significant SNP and time using \texttt{cox.zph} (Table~\ref{tab:sup_prop}).
Replication $p$-values in Table~\ref{tab:tab2} are one-sided $p$-values
for a logistic regression model controlling for age, sex, and five principal components.

For Figure~\ref{fig:survival}, we computed a myopia propensity score for each
individual as the (estimated) number of risk alleles among the 19 genome-wide
significant SNPs.  We then fit a Cox model including that score, sex, and
five principal components.  To estimate proportion variance explained for this
model, we used a pseudo-$r^2$ using likelihoods (similar to the Nagelkerke
pseudo $r^2$ for logistic regression).  That is, we calculated the variance
explained as 
\[
r^2 = \frac{1 - \frac{L} {nL_0}} {1 - \frac{1}{nL_0}},
\]
where $L_0$ is the null likelihood and $L$ the likelihood for the full model.
This is one of several methods used to compute variance explained for Cox
proportional hazards models \cite{pmid8896135}.

\section*{Acknowledgments}
We thank the customers of 23andMe for participating in this research and all
the employees of 23andMe for their contributions to this work.

\bibliography{myopia}

\onecolumn

\begin{appendix}

\setcounter{section}{19}  
\setcounter{figure}{0}
\setcounter{table}{0}
\renewcommand{\thefigure}{\thesection\arabic{figure}}
\renewcommand{\thetable}{\thesection\arabic{table}}

\section*{Supporting Information}

\begin{figure}[!ht]
\centering
\subfloat[][\gene{LAMA2} ]{\includegraphics[width=.48\textwidth]{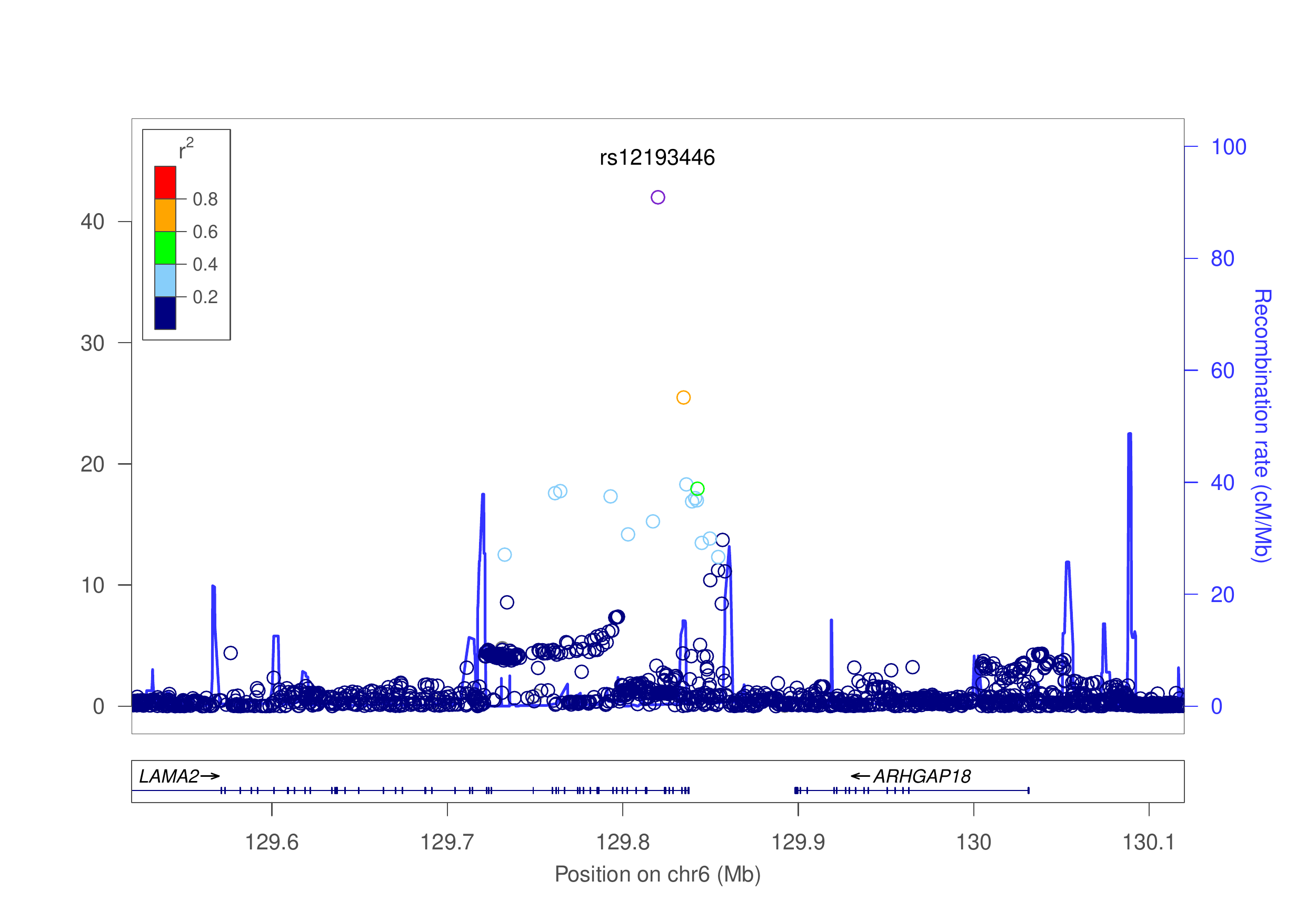}}
\subfloat[][\gene{LRRC4C}]{\includegraphics[width=.48\textwidth]{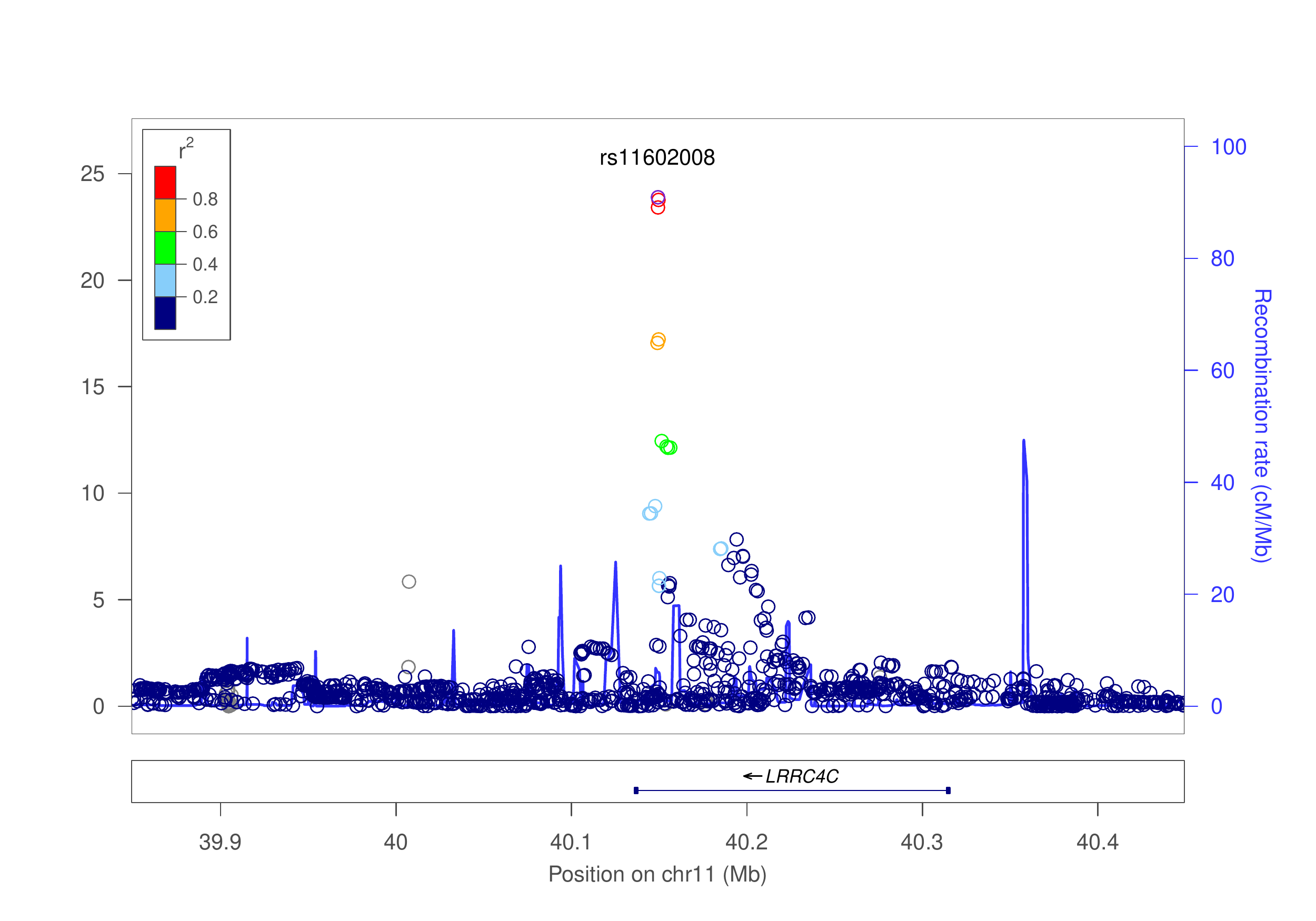}}\\
\subfloat[][\gene{RBFOX1}]{\includegraphics[width=.48\textwidth]{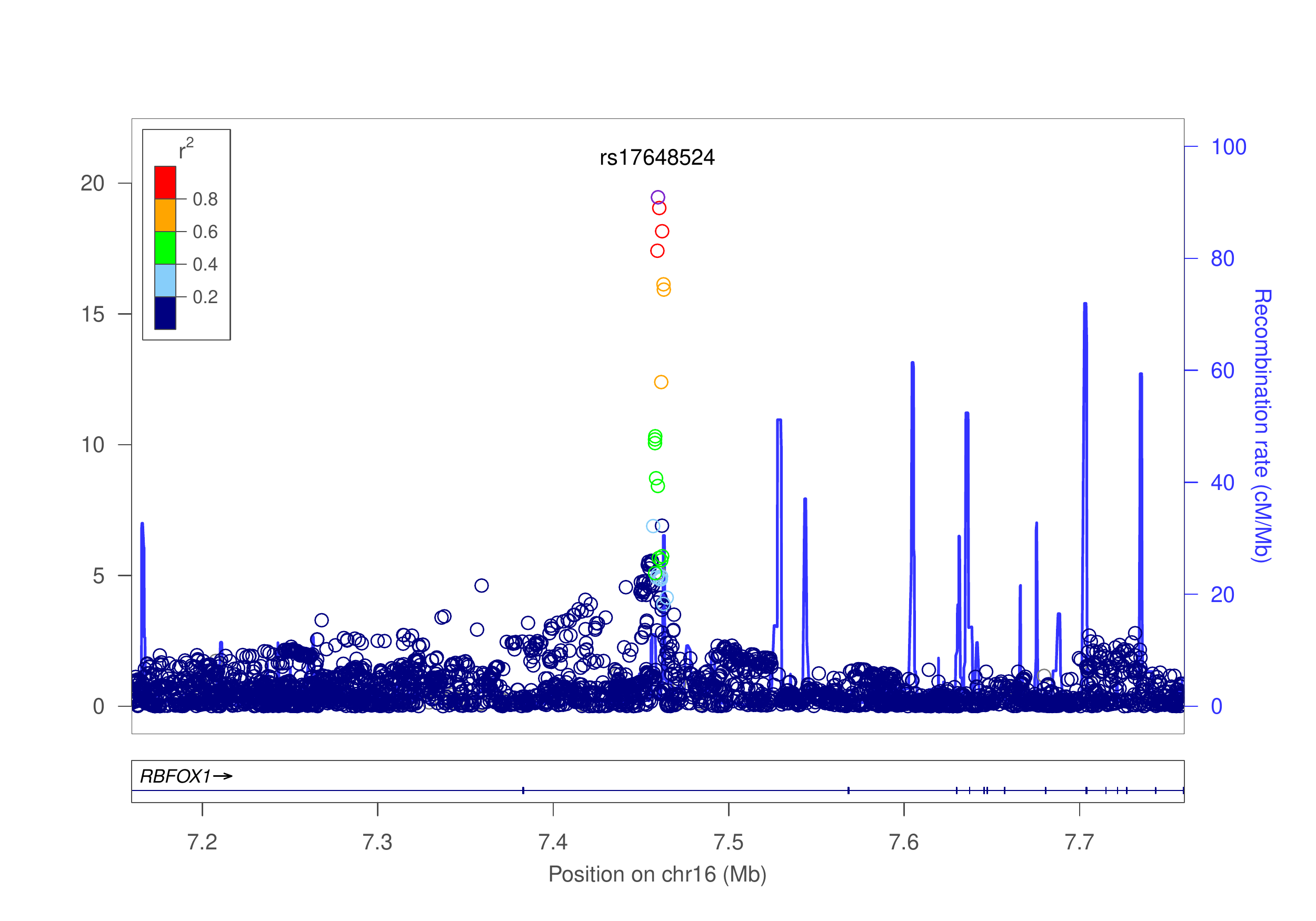}}
\subfloat[][\gene{RDH5}  ]{\includegraphics[width=.48\textwidth]{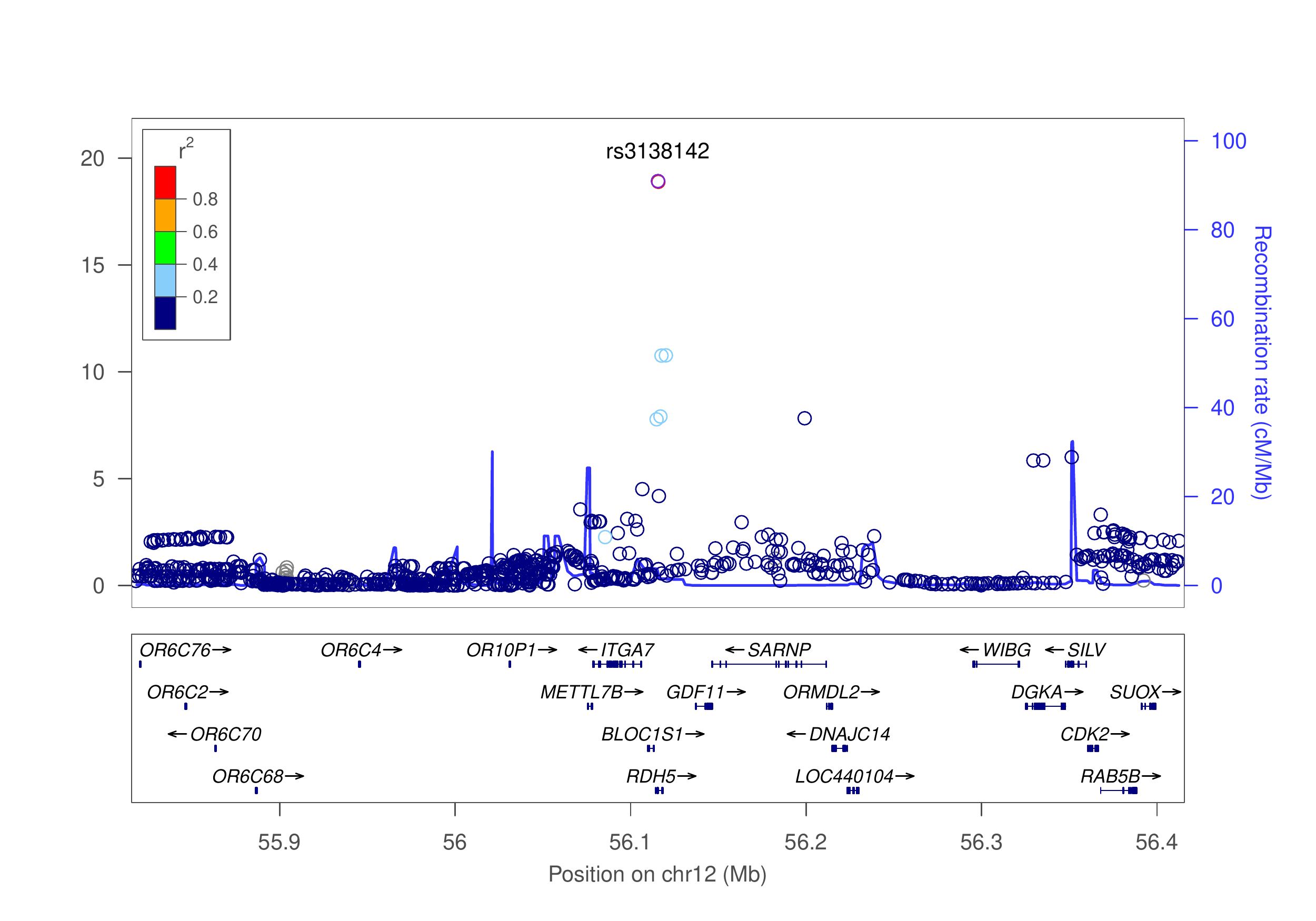}}
\caption{ {\bf Region plots for genome-wide significant associations} 
Colors depict the squared correlation ($r^2$) of each SNP with the most
associated SNP (shown in purple). Gray indicates SNPs for which $r^2$
information was missing.
}
\label{fig:regions}
\end{figure}

\begin{figure}[!ht]
\ContinuedFloat
\centering
\subfloat[][\gene{TOX/CA8}     ]{\includegraphics[width=.48\textwidth]{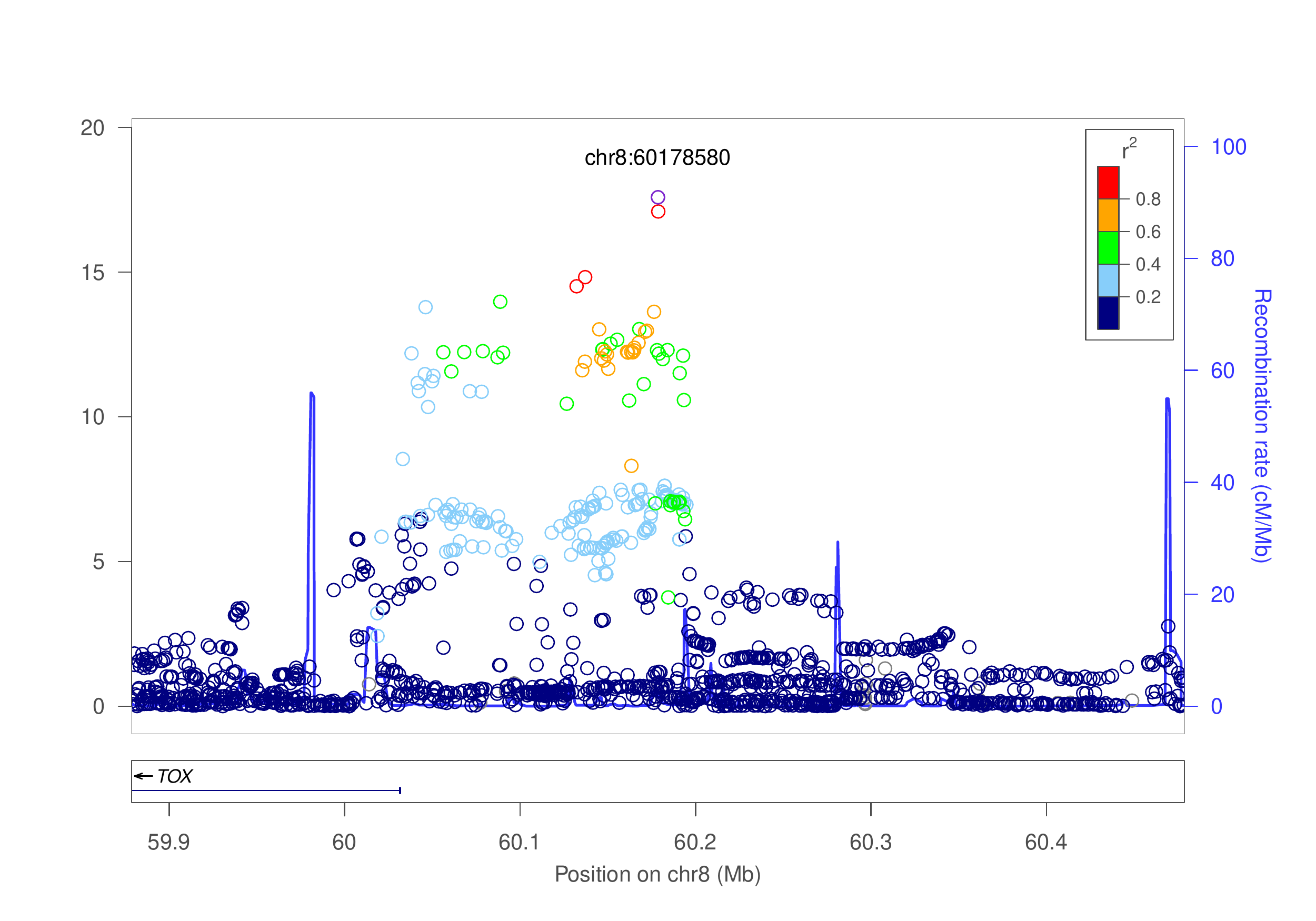}}
\subfloat[][\gene{KCNQ5}       ]{\includegraphics[width=.48\textwidth]{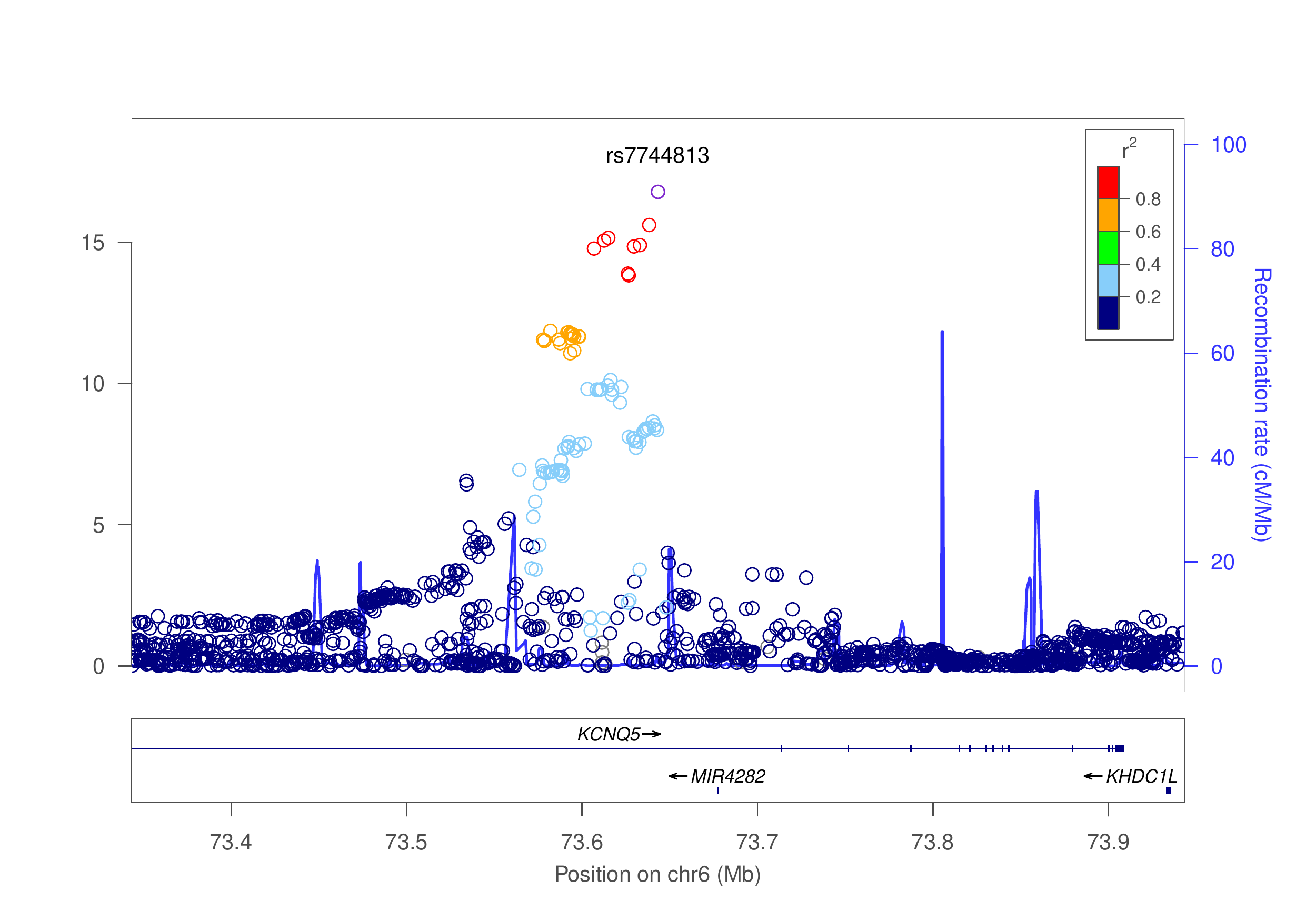}}\\
\subfloat[][\gene{GOLGA8B/GJD2}]{\includegraphics[width=.48\textwidth]{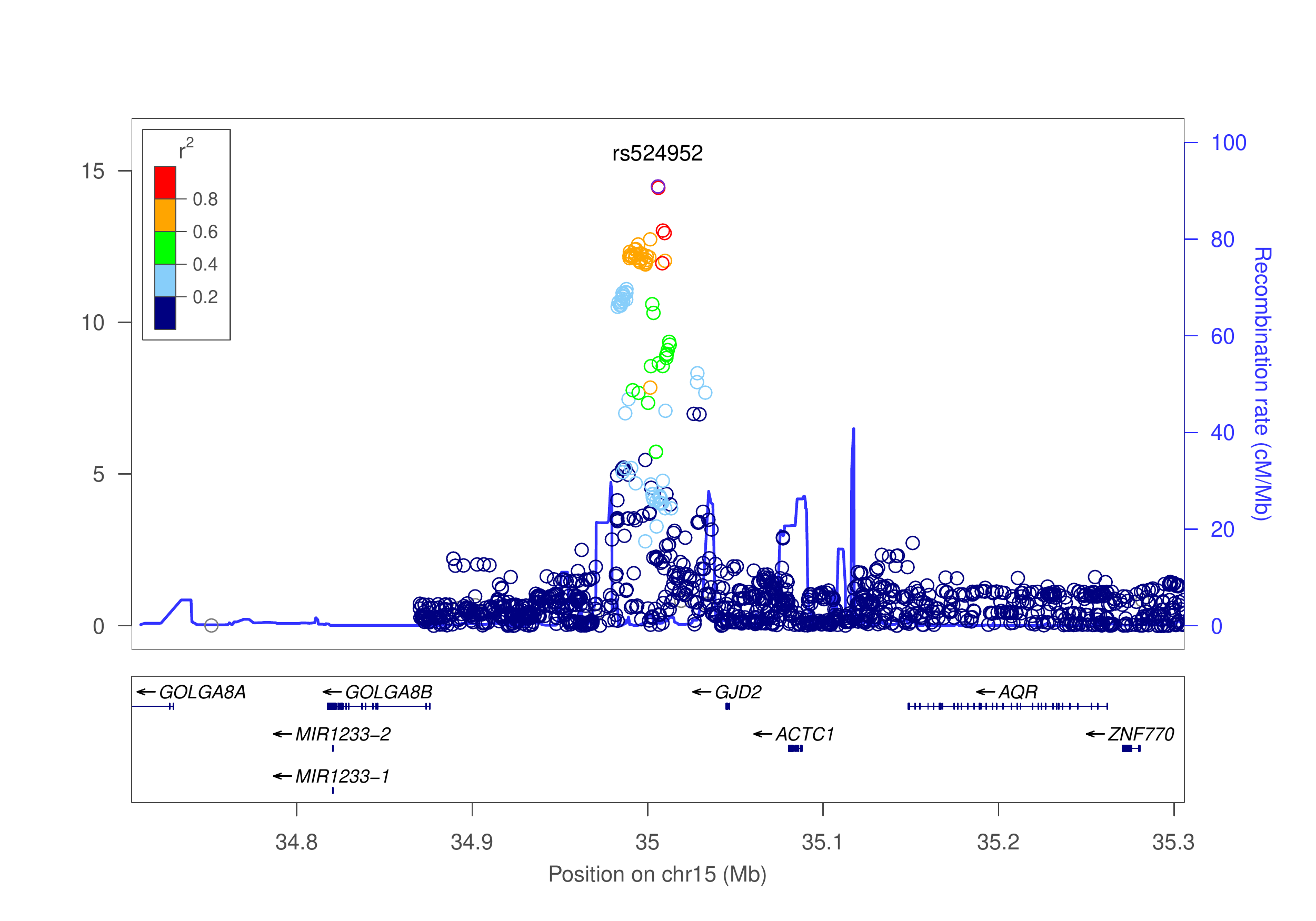}}
\subfloat[][\gene{SFRP1}       ]{\includegraphics[width=.48\textwidth]{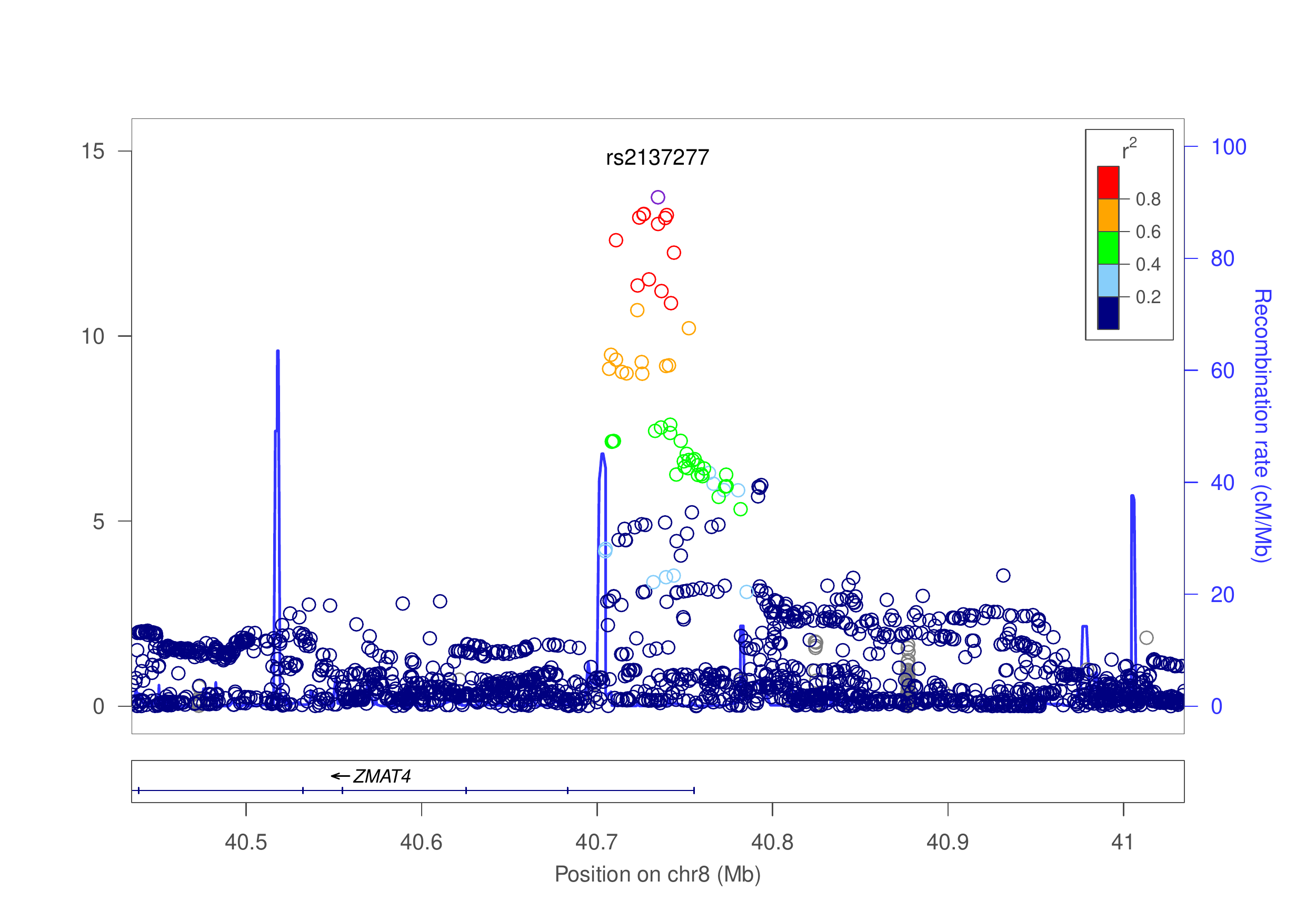}}
\caption{ {\bf Region plots for genome-wide significant associations} 
Colors depict the squared correlation ($r^2$) of each SNP with the most
associated SNP (shown in purple). Gray indicates SNPs for which $r^2$
information was missing.
}
\label{fig:regions}
\end{figure}

\begin{figure}[!ht]
\ContinuedFloat
\centering
\subfloat[][\gene{PRSS56} ]{\includegraphics[width=.48\textwidth]{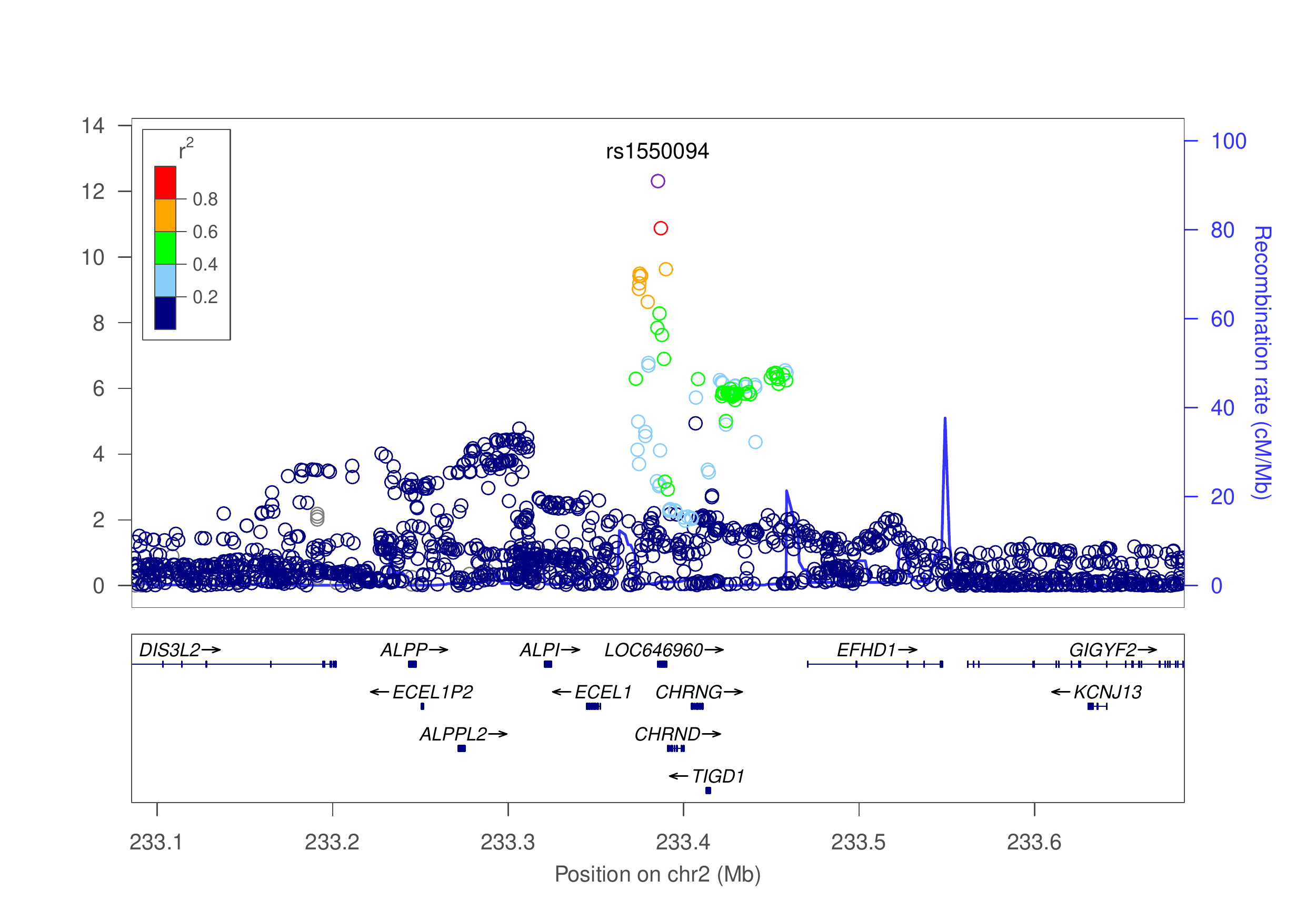}}
\subfloat[][\gene{PABPCP2}]{\includegraphics[width=.48\textwidth]{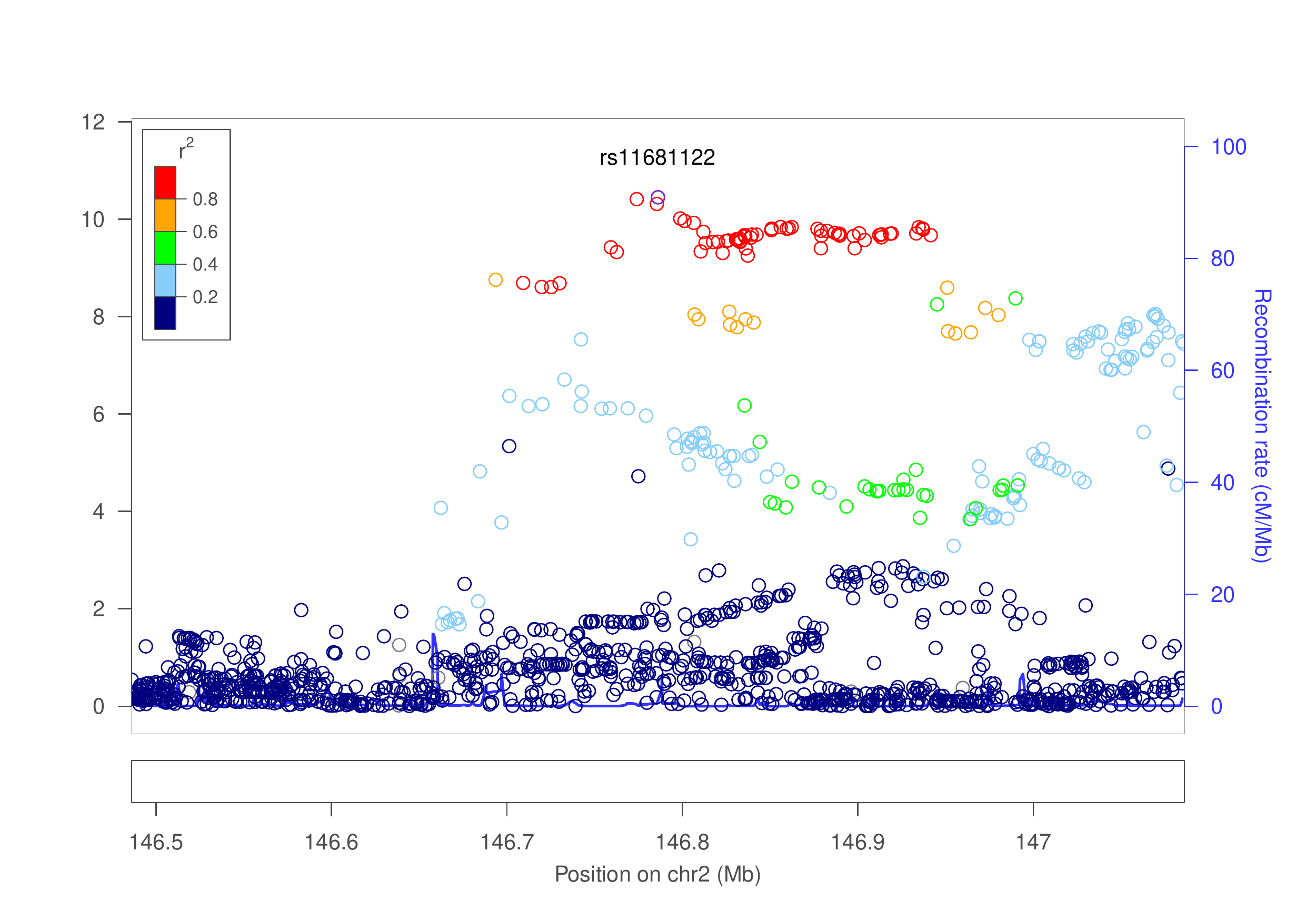}}\\
\subfloat[][\gene{ZBTB38} ]{\includegraphics[width=.48\textwidth]{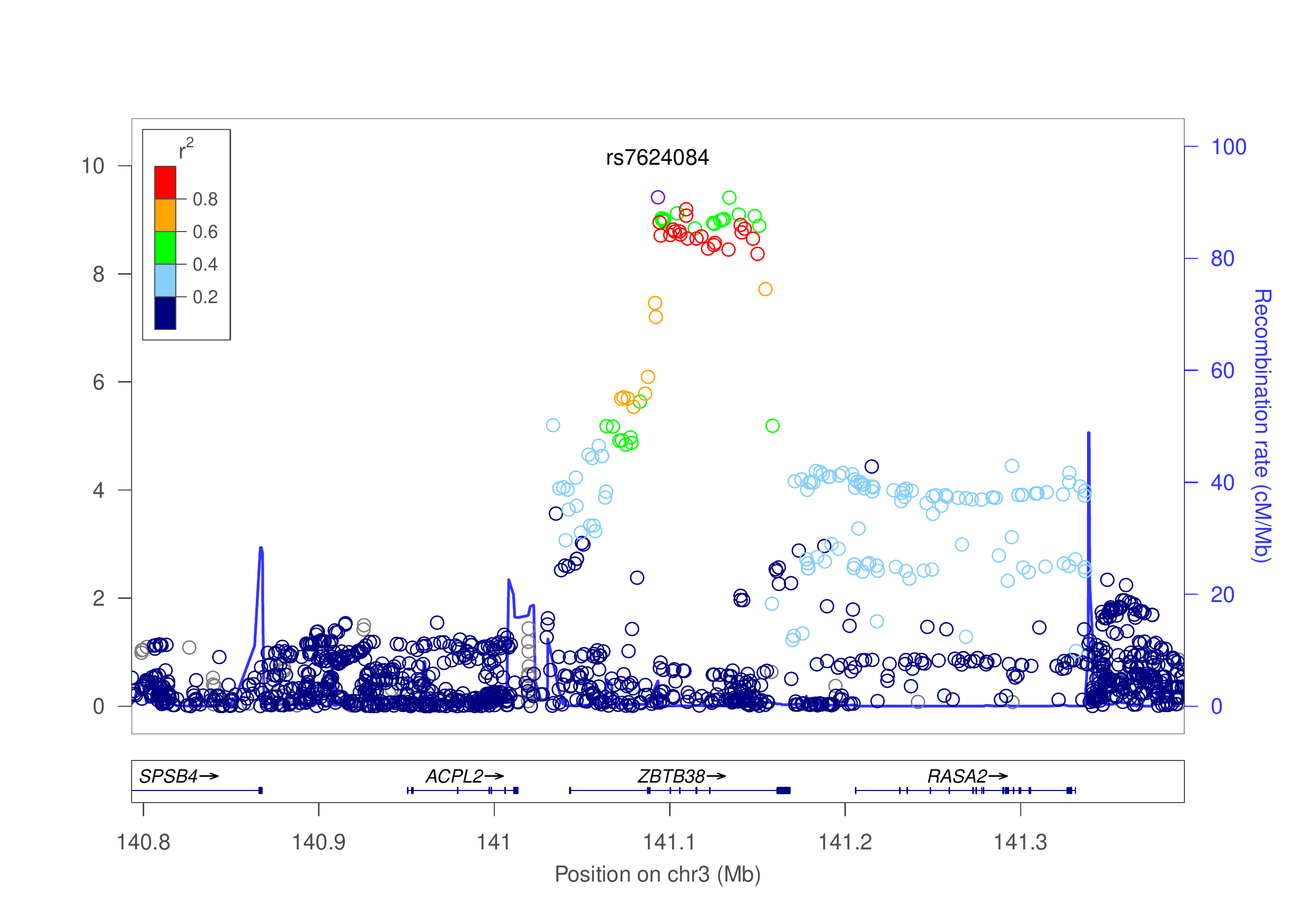}}
\subfloat[][\gene{PDE11A} ]{\includegraphics[width=.48\textwidth]{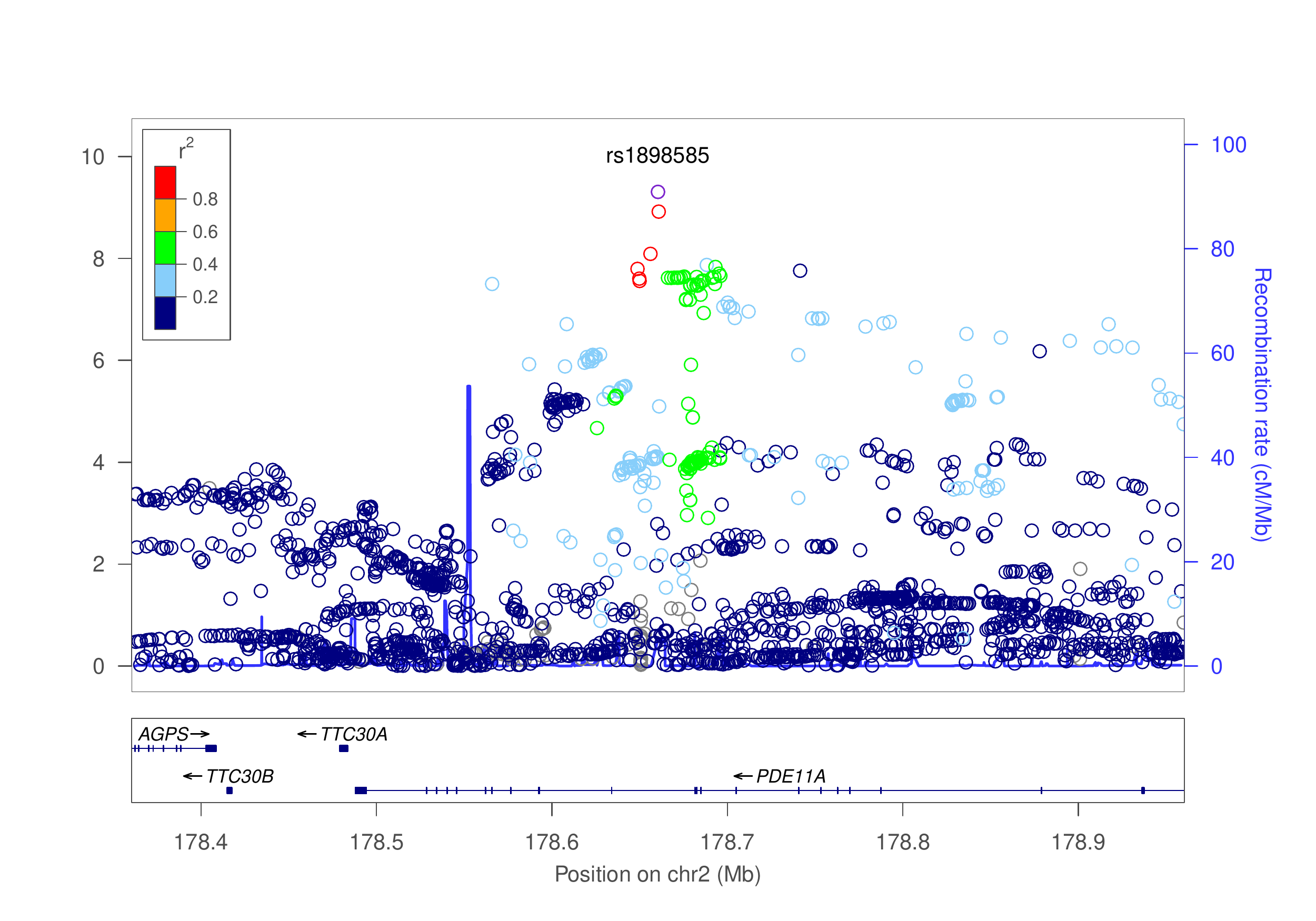}}
\caption{ {\bf Region plots for genome-wide significant associations} 
Colors depict the squared correlation ($r^2$) of each SNP with the most
associated SNP (shown in purple). Gray indicates SNPs for which $r^2$
information was missing.
}
\label{fig:regions}
\end{figure}

\begin{figure}[!ht]
\ContinuedFloat
\centering
\subfloat[][\gene{SHISA6}]{\includegraphics[width=.48\textwidth]{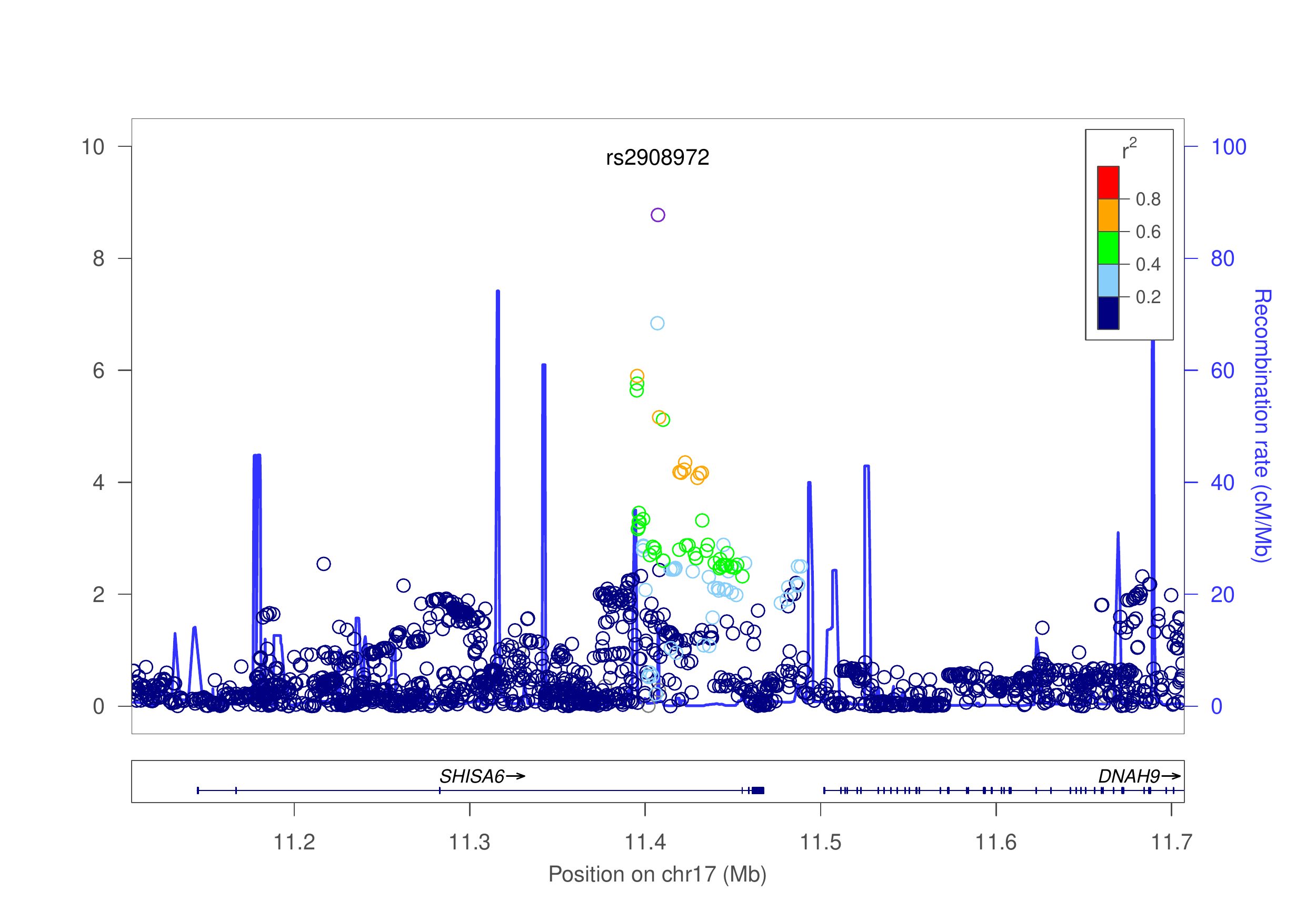}}
\subfloat[][\gene{KCNMA1}]{\includegraphics[width=.48\textwidth]{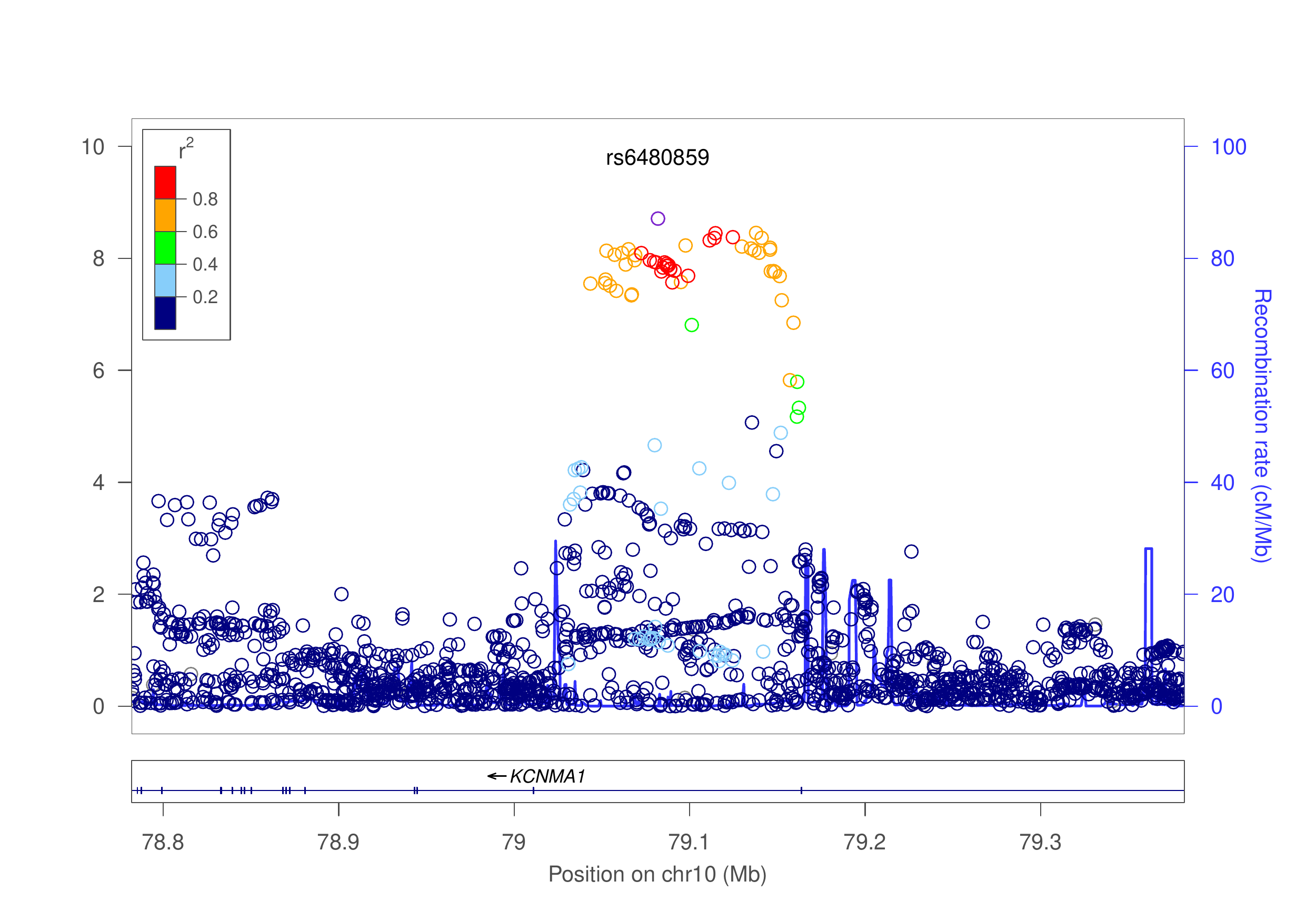}}\\
\subfloat[][\gene{DLG2}  ]{\includegraphics[width=.48\textwidth]{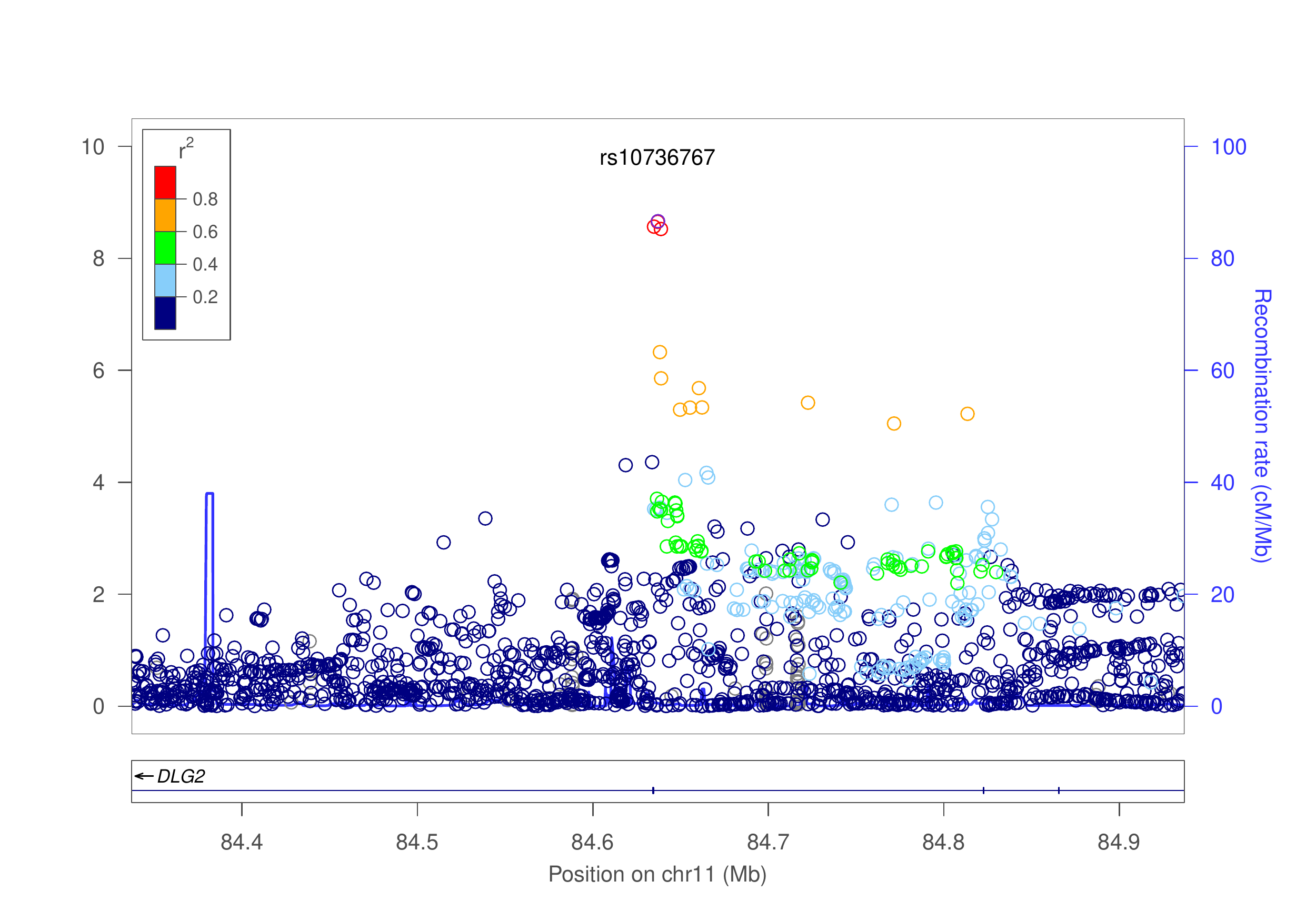}}
\subfloat[][\gene{TJP2}  ]{\includegraphics[width=.48\textwidth]{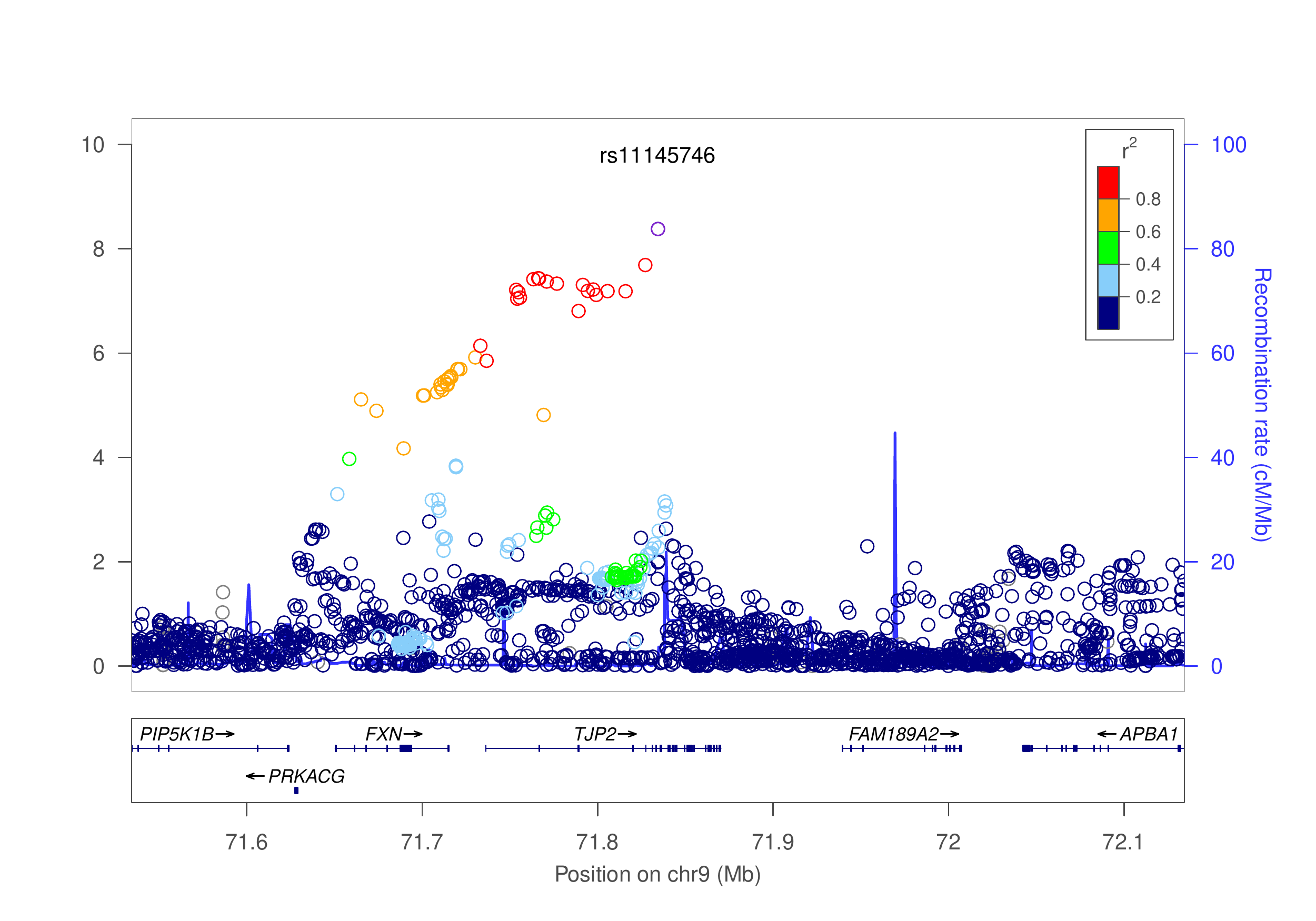}}
\caption{ {\bf Region plots for genome-wide significant associations} 
Colors depict the squared correlation ($r^2$) of each SNP with the most
associated SNP (shown in purple). Gray indicates SNPs for which $r^2$
information was missing.
}
\label{fig:regions}
\end{figure}

\begin{figure}[!ht]
\ContinuedFloat
\centering
\subfloat[][\gene{ZIC2}   ]{\includegraphics[width=.48\textwidth]{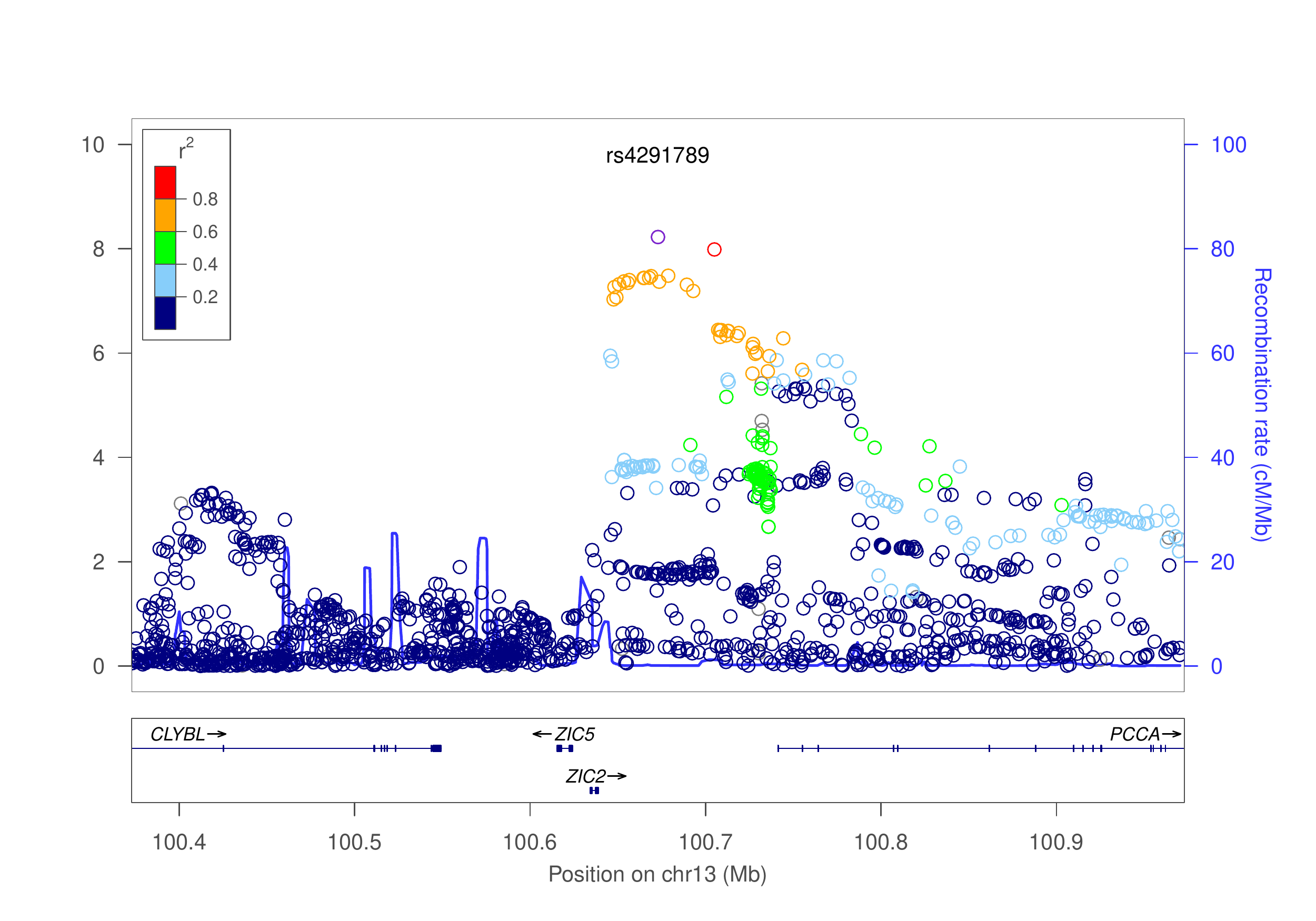}}
\subfloat[][\gene{RASGRF1}]{\includegraphics[width=.48\textwidth]{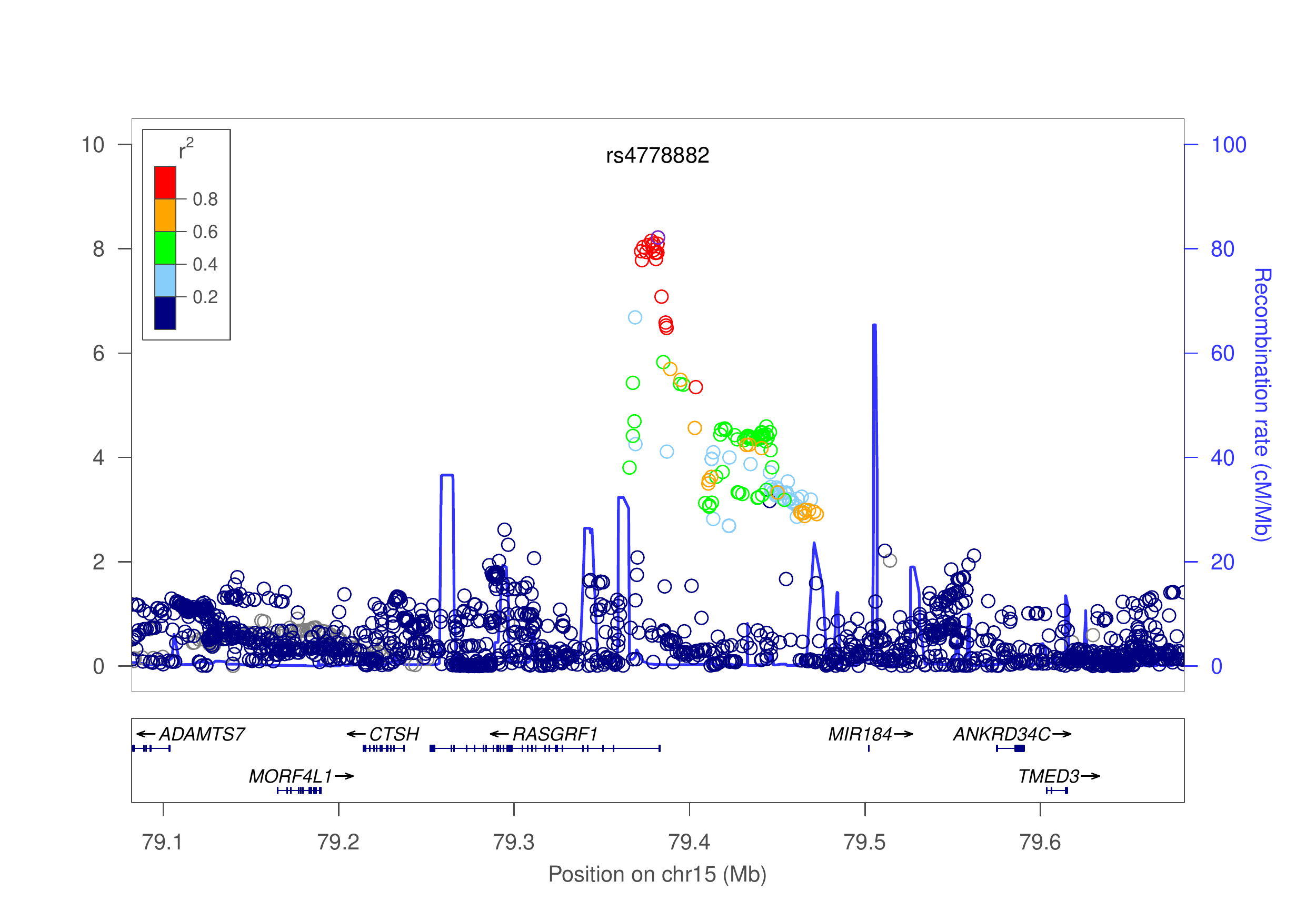}}\\
\subfloat[][\gene{RGR}    ]{\includegraphics[width=.48\textwidth]{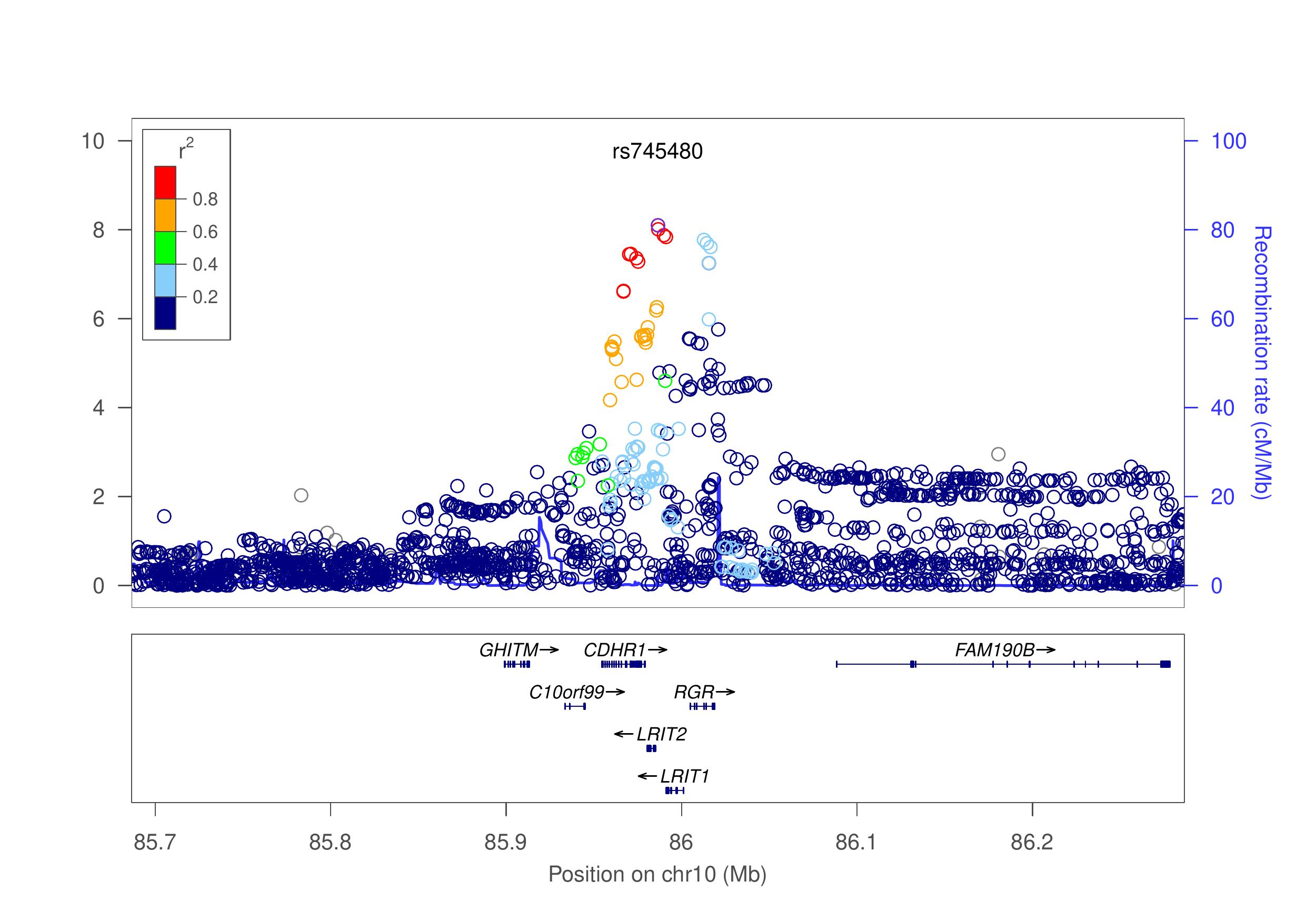}}
\caption{ {\bf Region plots for genome-wide significant associations} 
Colors depict the squared correlation ($r^2$) of each SNP with the most
associated SNP (shown in purple). Gray indicates SNPs for which $r^2$
information was missing.
}
\label{fig:regions}
\end{figure}

\begin{figure}[!ht]
\begin{center}
\includegraphics[width=4in]{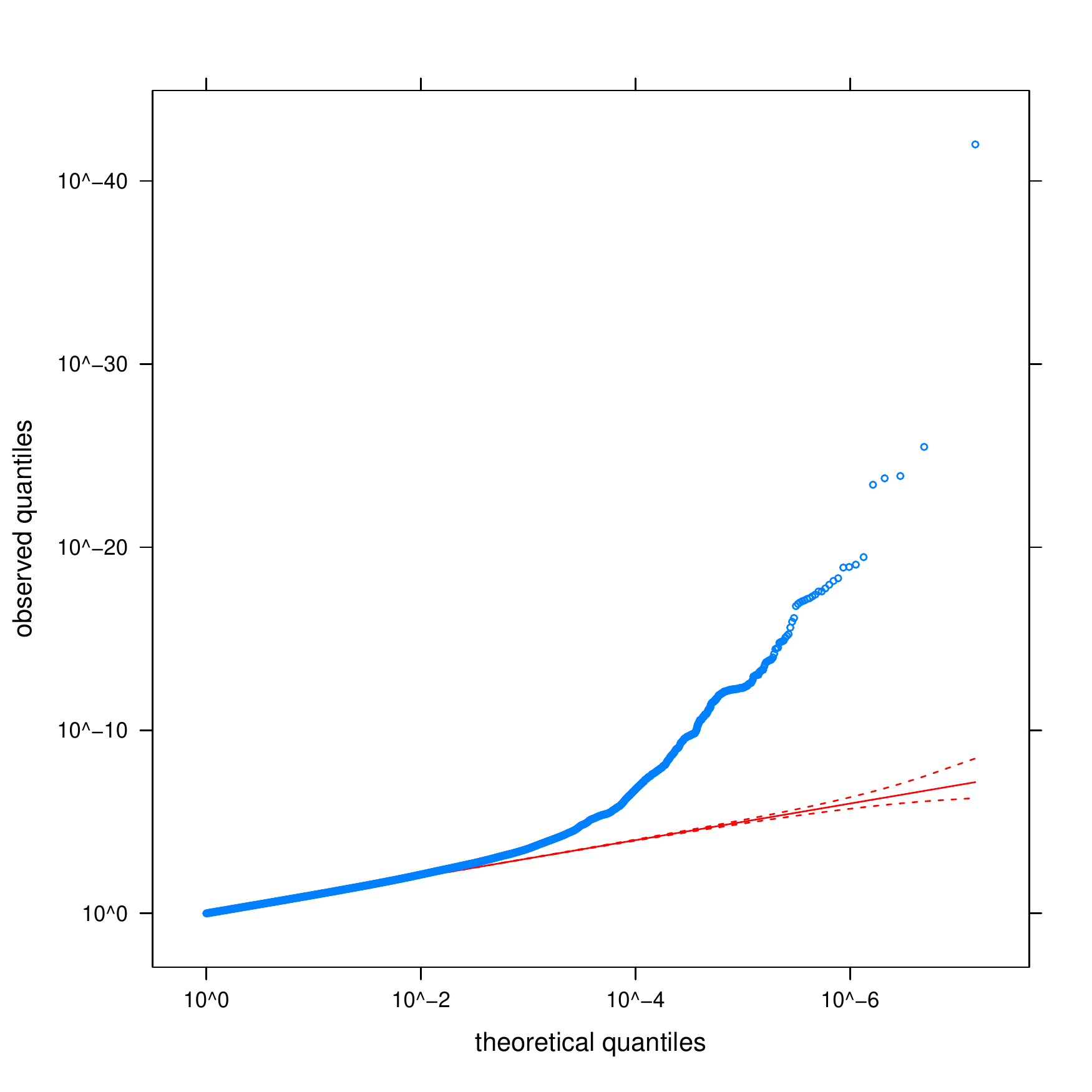}
\end{center}
\caption{ {\bf Quantile-quantile plot for myopia survival analysis} Actual ($\lambda$-corrected) $p$-values versus the null. }
\label{fig:qqplot}
\end{figure}


\begin{figure}[!ht]
\centering
\subfloat[][rs12193446, $p=5.5\cdot 10^{-9}$]{\includegraphics[width=.48\textwidth]{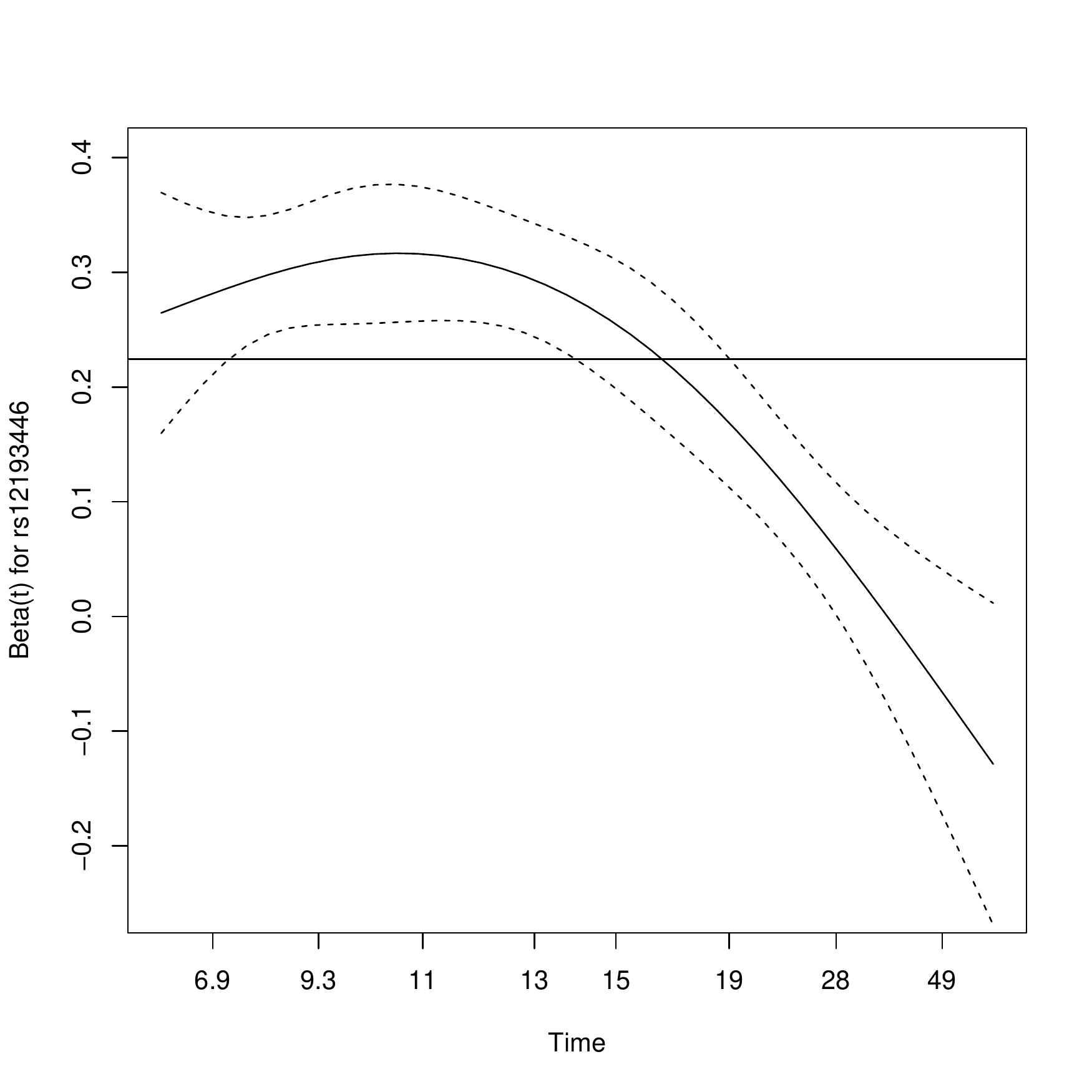}}
\subfloat[][rs17648524, $p=0.0011$]{\includegraphics[width=.48\textwidth]{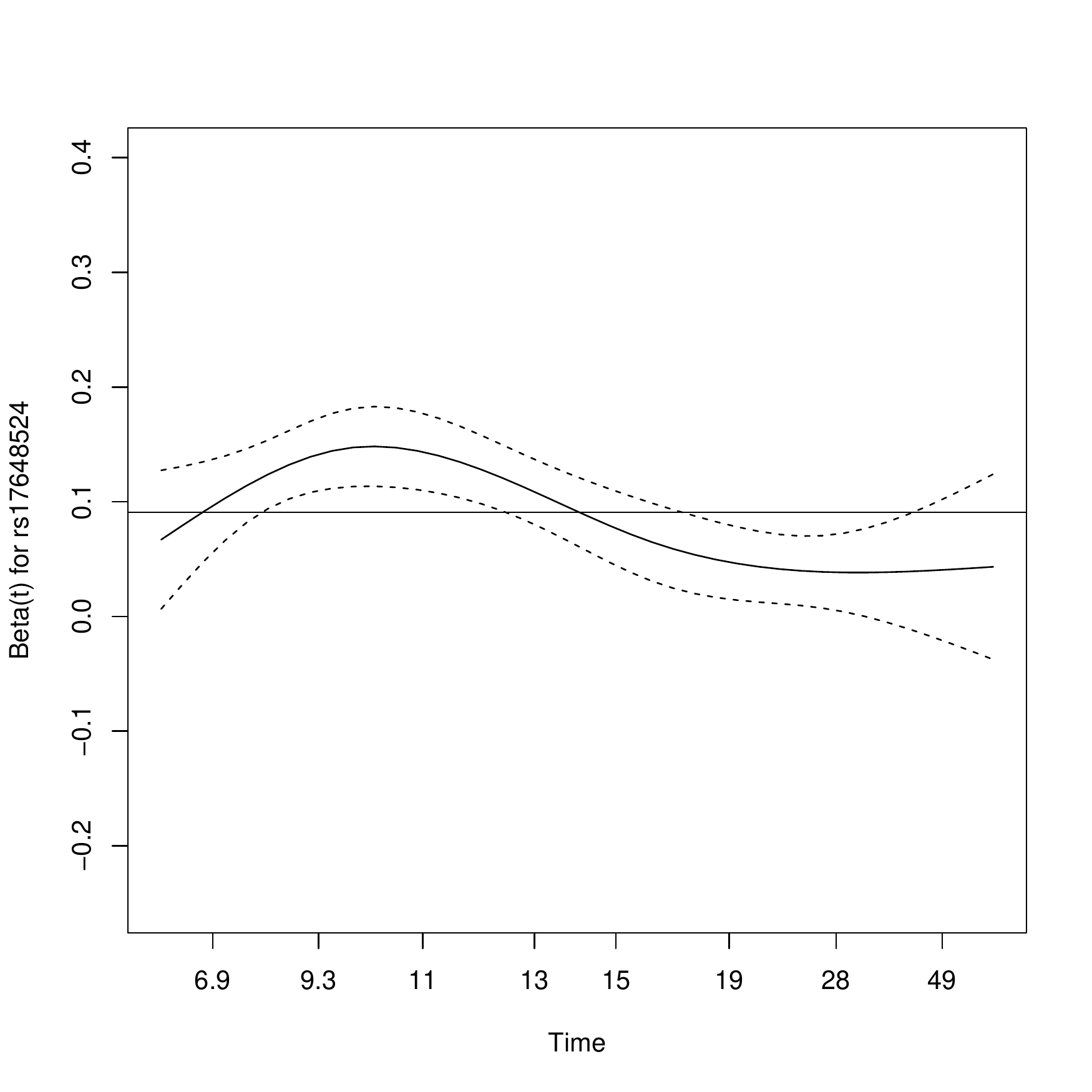}}\\
\subfloat[][rs3138142, $p=0.97$]{\includegraphics[width=.48\textwidth]{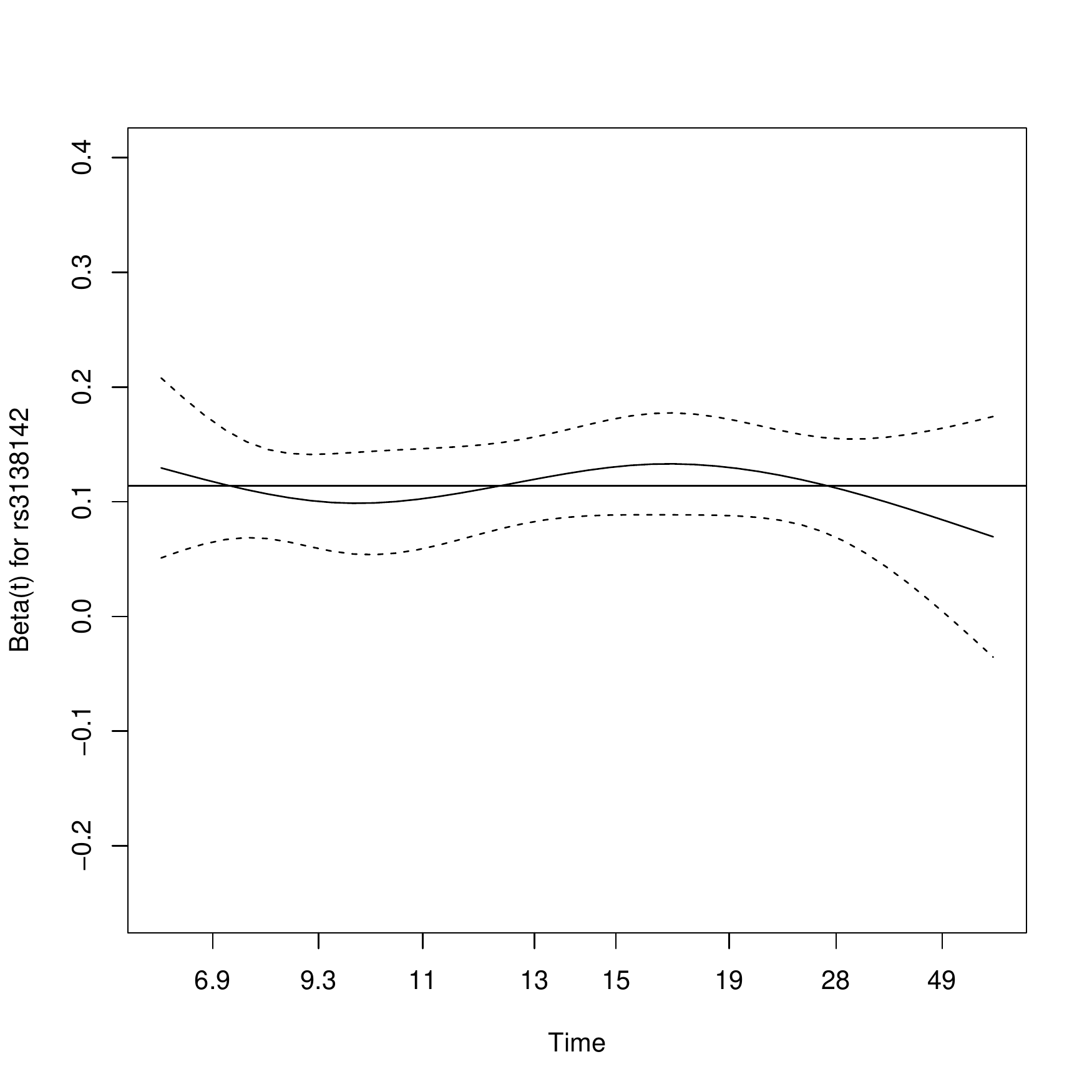}}
\subfloat[][chr8:60178580, $p=2.4\cdot 10^{-4}$]{\includegraphics[width=.48\textwidth]{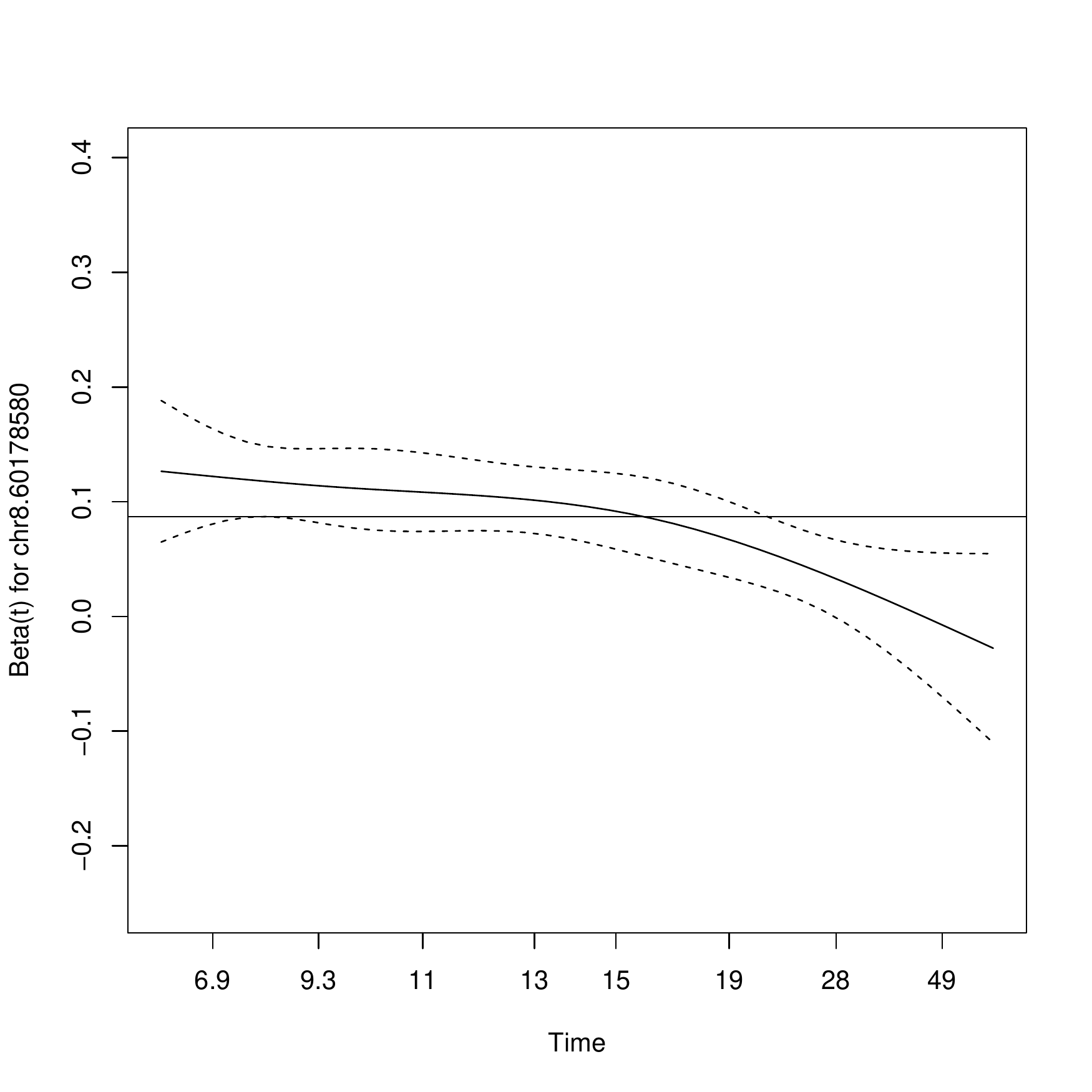}}
\caption{ {\bf Smoothed log-hazard ratios as a function of age for four representative SNPs} 
In each plot, the straight line shows the estimated log-hazard ratio (beta) for
each SNP in the proportional hazards model.  The solid curve is a spline fit to
beta estimated at different ages; the dotted curves are approximate 95\%
confidence intervals.  The $p$-value in each caption is the result of a test of
the proportional hazards assumption.  The sign of all coefficients has been
made positive for ease of comparison (so (a), (c), and (d) are flipped relative
to the main text). Note that among the examples here, only (c) shows no evidence of deviation from 
the proportional hazards assumption.
}
\label{fig:proportional}
\end{figure}

\begin{table}[!ht]
\caption{ \bf{$p$-values for survival and case-control analyses} }
\centering
\begin{tabular}{ccc}
\hline\hline
SNP & $p$ (survival) & $p$ (case-control)\\
\hline
   rs12193446 &  $  1\cdot10^{-42}$  &  $8.1\cdot10^{-33}$  \\
   rs11602008 &  $1.3\cdot10^{-24}$  &  $1.2\cdot10^{-20}$  \\
   rs17648524 &  $3.5\cdot10^{-20}$  &  $2.2\cdot10^{-17}$  \\
    rs3138142 &  $1.2\cdot10^{-19}$  &  $7.9\cdot10^{-20}$  \\  
chr8:60178580 &  $2.6\cdot10^{-18}$  &  $2.1\cdot10^{-14}$  \\
    rs7744813 &  $1.7\cdot10^{-17}$  &  $3\cdot10^{-15}$  \\
     rs524952 &  $3.3\cdot10^{-15}$  &  $8.8\cdot10^{-12}$  \\
    rs2137277 &  $1.8\cdot10^{-14}$  &  $4.8\cdot10^{-11}$  \\
    rs1550094 &  $4.9\cdot10^{-13}$  &  $1.7\cdot10^{-10}$  \\
   rs11681122 &  $3.6\cdot10^{-11}$  &  $1.6\cdot10^{-10}$  \\
    rs7624084 &  $3.8\cdot10^{-10}$  &  $1.3\cdot10^{-8}$  \\
    rs1898585 &  $4.9\cdot10^{-10}$  &  $7.9\cdot10^{-8}$  \\
    rs2908972 &  $1.7\cdot10^{-9}$  &  $8.3\cdot10^{-8}$  \\
    rs6480859 &  $2.0\cdot10^{-9}$  &  $7.9\cdot10^{-8}$  \\
   rs10736767 &  $2.2\cdot10^{-9}$  &  $1.6\cdot10^{-8}$  \\
   rs11145746 &  $4.2\cdot10^{-9}$  &  $3.1\cdot10^{-7}$  \\
    rs4291789 &  $  6\cdot10^{-9}$  &  $6\cdot10^{-7}$  \\
    rs4778882 &  $6.1\cdot10^{-9}$  &  $3\cdot10^{-8}$  \\
     rs745480 &  $  8\cdot10^{-9}$  &  $8.6\cdot10^{-8}$  \\
    rs5022942 &  $5.9\cdot10^{-8}$  &  $1\cdot10^{-6}$  \\
    rs9365619 &  $  1\cdot10^{-7}$  &  $2.9\cdot10^{-7}$  \\
    rs1031004 &  $1.5\cdot10^{-7}$  &  $1.7\cdot10^{-7}$  \\
   rs17428076 &  $1.6\cdot10^{-7}$  &  $2.8\cdot10^{-6}$  \\
chr14:54413001&  $4.6\cdot10^{-7}$  &  $2.8\cdot10^{-6}$  \\
     rs7272323&  $  7\cdot10^{-7}$  &  $1.2\cdot10^{-5}$  \\
chr11:65348347&  $7.9\cdot10^{-7}$  &  $6.2\cdot10^{-7}$  \\
    rs55819392&  $9.2\cdot10^{-7}$  &  $1.7\cdot10^{-6}$  \\
\hline\hline
\end{tabular}
\begin{flushleft}
$p$-values for SNPs in the survival analysis used in the paper as well as in a
case-control logistic regression on the same set of individuals.  The survival
analysis gives a smaller $p$-value for all but 1 SNP (rs3138142) and has 19
genome-wide significant ($p<5\cdot10^{-8}$) as compared to 13 for the
case-control. $p$-values in both cases are adjusted for the genomic control
inflation factor of 1.14.
\end{flushleft}
\label{tab:sup1}
\end{table}

\begin{table}[!ht]
\caption{ \bf{Tests of deviation from the proportional hazards assumption}}
\centering
\begin{tabular}{cc}
\hline\hline
SNP & $p$-value \\
\hline
   \textbf{rs12193446} &  $5.5\cdot 10^{-9}$ \\ 
   rs11602008          &     0.64 \\
   \textbf{rs17648524} &   0.0011 \\ 
    rs3138142          &     0.97 \\
\textbf{chr8:60178580} &  $2.4\cdot 10^{-4}$\\ 
    rs7744813          &    0.063 \\
     \textbf{rs524952} &  $3.9\cdot 10^{-4}$ \\ 
    rs2137277          &    0.016 \\
    \textbf{rs1550094} &   0.0026 \\ 
   rs11681122          &     0.59 \\
    rs7624084          &     0.32 \\
    rs1898585          &   0.0059 \\
    rs2908972          &    0.013 \\
    rs6480859          &     0.10 \\
   rs10736767          &    0.093 \\
   rs11145746          &   0.0038 \\
    rs4291789          &     0.37 \\
    rs4778882          &     0.34 \\
     rs745480          &    0.035 \\
\hline\hline
\end{tabular}
\begin{flushleft}
$p$-values for significant SNPs for deviation from the proportional hazards
assumption in the Cox model. For each SNP, we fit a Cox proportional hazards model
including the SNP, sex, and five principal components as predictors, and then tested
for independence of the scaled Schoenfeld residuals with time.
Bold SNPs deviate significantly
from this assumption after correction for 19 tests.  Plots for four example SNPs
are shown in Figure~\ref{fig:proportional}.
\end{flushleft}
\label{tab:sup_prop}
\end{table}

\end{appendix}
\end{document}